\documentclass[preprint,a4paper, number,times,sort&compress,12pt]{elsarticle} %one column
\journal{Computer Physics Communications}

\usepackage{graphicx}
\usepackage{rotating}
\usepackage{amsmath}
\usepackage{lscape}
\usepackage{color}
\usepackage{tabularx}
\usepackage{caption}
\usepackage{epsfig,color}
\usepackage{verbatim}
\usepackage{epstopdf}

\newcommand{\ai}{{\it ab initio}}
\newcommand{\cm}{cm$^{-1}$}

\newcommand{\p}{^\prime}

\newcommand{\duo}{\textsc{Duo}}
\newcommand{\Duo}{\textsc{Duo}}

\newenvironment{myverbatim}
{\endgraf\scriptsize\verbatim}
{\endverbatim}

\begin{document}

\begin{frontmatter}

\title{{\sc Duo}: a general program for calculating spectra of diatomic molecules \tnoteref{t1}}
\tnotetext[t1]{This paper and its associated computer program are available via the Computer Physics Communication homepage on ScienceDirect}

\author[ucl]{Sergei N. Yurchenko\corref{cor1}}
\ead{s.yurchenko@ucl.ac.uk}

\author[ucl]{Lorenzo Lodi}
\ead{l.lodi@ucl.ac.uk}

\author[ucl]{Jonathan Tennyson}
\ead{j.tennyson@ucl.ac.uk}

\author[moscow]{Andrey V. Stolyarov}
\ead{avstol@phys.chem.msu.ru}

\address[ucl]{Department of Physics \& Astronomy, University College London, Gower Street, London WC1E~6BT, United Kingdom}
\address[moscow]{Department of Chemistry, Lomonosov Moscow State University, Leninskiye gory 1/3, 119992 Moscow, Russia}

\cortext[cor1]{Corresponding author}

\begin{abstract}

{\sc Duo} is a general, user-friendly program for computing rotational,
rovibrational and rovibronic spectra of diatomic molecules. {\sc Duo}
solves the Schr\"{o}dinger equation for the motion of the nuclei not only
for the simple case of uncoupled, isolated electronic states (typical for
the ground state of closed-shell diatomics) but also for the general case
of an arbitrary number and type of couplings between electronic states
(typical for open-shell diatomics and excited states). Possible couplings
include spin-orbit, angular momenta, spin-rotational and spin-spin.
Corrections due to non-adiabatic effects can be accounted for by
introducing the relevant couplings using so-called Born-Oppenheimer
breakdown curves.

{\sc Duo} requires user-specified potential energy curves and, if
relevant, dipole moment, coupling and correction curves. From these it
computes energy levels, line positions and line intensities. Several
analytic forms plus interpolation and extrapolation options are available for representation of the
curves. {\sc Duo} can refine potential energy and coupling curves to best
reproduce reference data such as experimental energy levels or line
positions. {\sc Duo} is provided as a Fortran~2003 program and has been
tested under a variety of operating systems.

% Finally, {\sc Duo}  may also be used as a general-purpose solver for the time-independent
% one-dimensional Schr{\"o}dinger equation and is able to compute expectation values of arbitrary
% operators.
\end{abstract}


\end{frontmatter}

\section*{Program summary}
\noindent\emph{Program title:} {\sc Duo} \\
\emph{Catalogue number:} \\
\emph{Program summary URL:} \\
\emph{Program obtainable from:} CPC Program Library, Queen's University, Belfast, N. Ireland  \\
\emph{Licensing provisions:} Standard CPC licence. \\ %, http://cpc.cs.qub.ac.uk/licence/licence.html \\
\emph{No. of lines in distributed program, including test data, etc.:} 160~049 \\
\emph{No. of bytes in distributed program, including test data, etc.:} 13~957~785 \\
\emph{Distribution format:} tar.gz \\
\emph{Programming language:} Fortran 2003.\\
\emph{Computer:} Any personal computer.\\
\emph{Operating system:} Linux, Windows, Mac~OS.\\
\emph{Has the code been vectorized or parallelized?:} Parallelized.\\
\emph{Memory required to execute:} case dependent, typically $<$~10~MB\\
\emph{Nature of physical problem:} Solving the Schr{\"o}dinger equation for the nuclear motion of a diatomic molecule with an arbitrary number and type of couplings between electronic states. \\
\emph{Solution method}: Solution of the uncoupled problem first, then
basis set truncation and solution of the coupled problem. A line list can be computed if a dipole moment
function is provided. The potential energy and other curves can be empirically refined by fitting to experimental
energies or frequencies, when provided.  \\
\emph{Restrictions on the complexity of the problem:} The current version
is restricted to bound states of the system. \\
\emph{Unusual features of the program:} User supplied curves for all objects (potential
energies, spin-orbit and other couplings, dipole moment etc)  as analytic functions or
tabulated on a grid is a program requirement.\\
\emph{Typical running time:} Case dependent. The test runs provided take seconds or a few minutes on a normal PC.\\

\section{Introduction}
\label{sec:intro}

Within the Born-Oppenheimer or adiabatic approximation \cite{Cederbaum2004} the
rotational-vibrational (rovibrational) energy levels of a diatomic molecule
with nuclei $a$ and $b$ and in a $^1\Sigma^\pm$ electronic state are given by
the solution of the one-dimensional Schr{\"o}dinger equation:
\begin{equation}\label{1d-SE}
-\frac{\hbar^2}{2 \mu} \frac{\mathrm{d}^2}{\mathrm{d}r^2}\psi_{\upsilon J}(r) + \left[V_{\rm state}(r) + \frac{J(J+1)}{2 \mu r^2} \right] \psi_{\upsilon J}(r) = E_{\upsilon J} \psi_{\upsilon J}(r),
\end{equation}
where $\mu^{-1} = {M_a}^{-1} + {M_b}^{-1}$ is the reduced mass of the
molecule and $M_a$ and $M_b$ are the (nuclear) masses of atoms $a$ and $b$,
respectively. $V_{\rm state}(r)$ is the potential energy curve (PEC) for the
electronic state under study, $J$ is the total angular momentum of the
molecule and $\upsilon=0, 1, \ldots$ is the vibrational quantum number. The
solution of this one-dimensional Schr{\"o}dinger equation is a
well-studied mathematical problem \cite{Berezin.Shubin, Simon2000} for which
many efficient numerical methods are available \cite{Blatt1967, Shore1973,
Johnson1977, Johnson1978, Korsch1981, Guardiola1982, Guardiola1982a,
Stolyarov1987, Lindberg1988, Marston1989, Abarenov1990, Fargas1996,
Fargas1997, Ishikawa2002, Utsumi2004, Wang2004}; the most popular of them is
probably the iterative ``shooting'' Cooley-Numerov
\cite{61Coxxxx.method,Cashion1963,24Nuxxxx.method} method which is notably used in the
program {\sc Level} due to Le Roy \cite{lr07}.

As well as the `direct' problem of solving the Schr{\"o}dinger equation for a
given PEC, also of great interest is the corresponding inverse problem
\cite{Karkowski2009,Weymuth2014}, that is the task of determining the
potential $V_{\rm state}(r)$ which leads to a given set of energy levels
$E_{\upsilon J}$, typically obtained from experiment. A traditional way of
performing this task approximately is to use the semi-classical
Rydberg-Klein-Rees (RKR) method \cite{Karkowski2009}; a more precise strategy
called inverse
perturbation analysis (IPA) has been suggested %based on perturbation theory
by Kosman and Hinze \cite{Kosman1975,Weymuth2014} and a program implementing
this approach was presented by Pashov \emph{et al} \cite{Pashov2000}.
A different, grid-based fitting strategy has been recently suggested by
Szidarovszky and Cs{\'a}sz{\'a}r \cite{14SzCsxx.methods}.
The program {\sc DPotFit} \cite{dpotfit}, a companion to
Le Roy's {\sc Level}, %provides a method of achieving this
also provides this functionality
for isolated states of closed shell diatomics.
Indeed, for single potential problems there is an extensive literature on the
determination of potential curves %, and Born-Oppenheimer breakdown (BOB) corrections,
from experimental data; in this context we particularly note the work of Coxon
and Hajigeorgiou \cite{92CoHa.CO,ch99,04CoHa.CO} and Le Roy and
co-workers \cite{04Lexxxx.NaCl, 11LeHaTa.method, 14MeStHe.method,
15WaSeLe.NaH}.

% Approaches which recover the potential Refs to direct potential fit methods (DPF). Indeed, for single potential problems there is an extensive literature of the determination potential curves, and Born-Oppenheimer breakdown (BOB)

When the diatomic molecule has a more complex electronic structure
(i.e., the electronic term is not $^1\Sigma$)
% with several coupling curves some of which are degenerate,
the situation is more
complicated, as interactions between the various terms are present and it is
not possible to treat each electronic state in isolation. Although there are a growing
number of studies treating coupled electronic states, for example see
Refs.~\cite{95CaLeMa.diatom,02TaFeZa.NaRb,03BeFeGu.RbCs,05MeZa.LiAr,
08HuTiJu.diatom,10ZhSaDa.diatom,13GoAbHa.diatom,14BrRsWe.CN}, there
appears to be no general program available for solving the coupled problem,
the closest being a general coupled-state program due to Hutson~\cite{94Hutson}.
We have therefore developed a new computer program, \Duo, particularly to
deal with such complex cases.

\Duo\ is a flexible, user-friendly program written in Fortran 2003 and
capable of solving both the direct and the inverse problem for a general
diatomic molecule with an arbitrary number and type of couplings between
electronic states, including spin-orbit, electronic-rotational,
spin-rotational and spin-spin couplings. \Duo\ also has auxiliary
capabilities such as interpolating and extrapolating curves and calculating
lists of line positions and line intensities (so-called line lists). \Duo\ is
currently being used as part of the ExoMol project \citep{jt528}, whose aim
is to generate high-temperature spectra for all molecules likely to be
observable in exoplanet atmospheres in the foreseeable future. Completed
studies based on the use of \Duo\ include ones on AlO \cite{jt589,jt598},
ScH \cite{jt599}, CaO \cite{jt618} and VO \cite{jt625}. Our methodology
is the subject of a very recent topical review \cite{jtdiat}.

This paper is organised as follows. In Section~\ref{s:duo} we review the
theory and the basic equations used by \Duo\ to solve the coupled nuclear
motion problem for diatomics. In Section~\ref{sec.intensities} we discuss the
calculation of molecular line intensities and line lists.
Section~\ref{s:refinement} is devoted to the inverse problem, i.e. to the
refinement (`fitting') of potential and coupling curves so that they
reproduce a set of reference energy levels or line positions. Section
\ref{s:functions} reviews the functional forms implemented for the various
curves. In Section~\ref{s:program} the program structure is explained.
Finally, we draw our conclusions in Section \ref{s:conclusion}. Technical
details on program usage such as detailed explanations of the program options
and sample inputs are reported in a separate user's manual.

\section{Method of solution} \label{s:duo}

% \subsection{Hamiltonian and matrix elements}
After separating out the centre-of-mass motion and having introduced a
body-fixed set of Cartesian axes with origin at the centre of nuclear mass
and with the $z$ axis along the internuclear
direction the non-relativistic Hamiltonian of a diatomic molecule
can be written as
\cite{Islampour2015,Sutcliffe2007,93Kato.methods,Pack1968,Bunker1968}:
\begin{equation}\label{e:H-tot}
\hat{H}_{\rm tot}=  \hat{H}_{\rm e} + \hat{H}_\mu + \hat{H}_{\rm vib} + \hat{H}_{\rm rot}
\end{equation}
where the meaning of the various terms is as follows. $\hat{H}_{\rm e}$  is the
electronic Hamiltonian and is given by
\begin{equation}\label{e:H-elec}
\hat{H}_{\rm e} =-\frac{\hbar^2}{2 m_e} \sum_{i=1}^{N_e} {\nabla_i}^2  + V(r,\xi_i)
\end{equation}
where $V(r,\xi_i)$ is the Coulomb electrostatic interactions between all particles
(electrons and nuclei) and we indicated with $r$ the internuclear coordinate and collectively
with $\xi_i$ the full set of electron coordinates; $\hat{H}_\mu$ is the mass-polarisation term given by
\begin{equation}\label{e:H-massP}
\hat{H}_\mu =-\frac{\hbar^2}{2 m_N} \sum_{i=1}^{N_e} \sum_{j=1}^{N_e} {\nabla_i} \cdot {\nabla_j}
\end{equation}
where $m_N$ is the total nuclear mass; $\hat{H}_{\rm vib} $ is
the vibrational kinetic energy operator and is given by
\begin{equation}\label{e:H-vib}
\hat{H}_{\rm vib} = -\frac{\hbar^2}{2 \mu} \frac{\mathrm{d}^2}{\mathrm{d}r^2}
\end{equation}
where $\mu$ is the reduced mass of the molecule.
$\hat{H}_{\rm rot}$ is the rotational Hamiltonian and can be expressed
in terms of the body-fixed rotational angular momentum (AM) operator as
\begin{equation}
\hat{H}_{\rm rot} = \frac{\hbar^2}{2 \mu r^2} \mathbf{\hat{R}}^2 .
\end{equation}
In turn, the rotational AM can be expressed as $\mathbf{\hat{R}} =
\mathbf{\hat{J}}-\mathbf{\hat{L}}-\mathbf{\hat{S}}$ where $\mathbf{\hat{J}}$
is the total AM, $\mathbf{\hat{L}}$ is the electron orbital AM and
$\mathbf{\hat{S}}$ is the electron spin AM. The total AM operator
$\mathbf{\hat{J}}$ acts on the Euler angles $(\theta,\phi,\chi)$ relating the
laboratory-fixed and the body-fixed Cartesian frame and its expression can be
found, e.g., in Ref.~\cite{Islampour2015}. Introducing the ladder operators
$\hat{J}_{\pm} = \hat{J}_x \pm i \hat{J}_y$, $\hat{S}_{\pm} = \hat{S}_x \pm i
\hat{S}_y$ and $\hat{L}_{\pm} = L_x \pm i L_y$  we can express the rotational
Hamiltonian as
\begin{eqnarray}
\nonumber
 \hat{H}_{\rm rot} & = & \frac{\hbar^2}{2 \mu r^2}   \left[ (\hat{J}^{2}-\hat{J}_{z}^{2})+ (\hat{L}^{2}-\hat{L}^{2}_{z}) +(\hat{S}^{2}-\hat{S}_{z}^2) + \right. \\
 \label{e:Hr}
 &+& \left.  (\hat{J}_{+}\hat{S}_{-}+\hat{J}_{-}\hat{S}_{+}) -(\hat{J}_{+}\hat{L}_{-}+\hat{J}_{-}\hat{L}_{+})+(\hat{S}_{+}\hat{L}_{-}+\hat{S}_{-}\hat{L}_{+})  \right].
\end{eqnarray}

The approach used by \Duo\ to solve the total rovibronic Schr{\"o}dinger equation
with the Hamiltonian (\ref{e:H-tot}) follows closely the standard
coupled-surface Born-Oppenheimer treatment \cite{Cederbaum2004,Kutzelnigg1997,Hutson1980}.
It is assumed that one has preliminary solved the electronic motion problem with clamped nuclei
\begin{equation}
\hat{H}_{\rm e} | \mathrm{state}, \Lambda, S, \Sigma \rangle = V_\mathrm{state}(r) | \mathrm{state}, \Lambda, S, \Sigma \rangle
\end{equation}
for all electronic states of interest.  The electronic wave functions depend
on the electron coordinates $\xi_i$ and parametrically on the internuclear
distance $r$ and can be labelled by total spin $S=0,1/2,1,\ldots$,
projection of $\mathbf{\hat{L}}$ along the body fixed $z$ axis $\Lambda=0,\pm
1, \pm 2$, projection of $\mathbf{\hat{S}}$ along the body fixed $z$ axis
$\Sigma=0,\pm 1/2,\pm 1,\ldots$ and by a further label `state'=$1,2,\ldots$
which counts over the electronic curves. For $|\Lambda |\geq 1$ the spacial
part of the electronic wave functions is doubly degenerate; we choose the
degenerate components $ | {\rm state}, \Lambda, S, \Sigma \rangle$ so that
they satisfy the following conditions \cite{93Kato.methods}:
\begin{eqnarray}
\label{e:<|Lz|>}
 \langle  {\rm state}, \Lambda, S, \Sigma  |\hat{L}_{z}| {\rm state}, \Lambda, S, \Sigma \rangle &=& \Lambda, \\
 \label{e:sigmav}
 \hat{\sigma}_{v}(xz)  | {\rm state}, \Lambda, S, \Sigma \rangle &=&  (-1)^{s-\Lambda+S-\Sigma} | {\rm state}, -\Lambda, S, -\Sigma \rangle,
\end{eqnarray}
where $\hat{\sigma}_{v}(xz) $ is the symmetry operator corresponding to
a reflection through the body-fixed $xz$-plane (parity operator)
and $s=1$ for $|\Sigma^{-} \rangle$ states and $s=0$ for all other states.

Once the potential energy curves $V_\mathrm{state}(r)$ have been obtained,
for example  using an \emph{ab initio} quantum chemistry program, \Duo\
solves the rotationless ($J=0$) one-dimensional Schr{\"o}dinger given by
Eq.~(\ref{1d-SE}) separately for each electronic curve $V_{\rm state}(r)$,
producing a set of vibrational eigenvalues $E_\upsilon$ and vibrational wave
functions $|{\rm state}, \upsilon \rangle$, where $\upsilon=0, 1, \ldots$ is
the vibrational quantum number assigned on the basis of the energy ordering;
technical details on this step are given in Section \ref{solution.SE}. A
subset of $N_\upsilon({\rm state})$
vibrational functions are selected to form a basis set of rovibronic %electronic-v{\rm state}ibrational-rotational
basis functions defined by
\begin{equation}\label{e:basis}
 |{\rm state}, J, \Omega, \Lambda, S, \Sigma, \upsilon \rangle  =  | {\rm state}, \Lambda, S, \Sigma  \rangle |{\rm state},\upsilon \rangle | J,\Omega,M \rangle ,
\end{equation}
where  $| J, \Omega,M \rangle$ is a symmetric-top eigenfunction \cite{Islampour2015} (a function of
the Euler angles) and describes the overall rotation of the molecule as a whole,
$\Omega= \Lambda+\Sigma$ and $M$ is the projection of the
total angular momentum along the laboratory axis $Z$.
Only combinations of $\Sigma$ and $\Lambda$ which satisfy $|\Omega|\le \min(J,|\Lambda|+S) $ are selected
in the rovibronic basis set (\ref{e:basis}).
The selection of vibrational basis functions to retain %for the $J>0$ and/or coupled problem
can be made either by specifying an
energy threshold (all vibrational states below the threshold are retained) or
by specifying a maximum vibrational quantum number $\upsilon_{\rm max}$.

The rovibrational basis set (\ref{e:basis}) is used to solve the complete
rovibronic Hamiltonian given by Eq.~(\ref{e:H-tot}); this amounts to using an
expansion in Hund's case (a) functions to solve the coupled problem. In
particular, the ladder operators appearing in $\hat{H}_{\rm rot}$ couple
rovibrational states belonging to different electronic states; specifically,
the non-vanishing matrix elements of the angular momentum operators in the
rotational Hamiltonian \eqref{e:Hr} are given by the standard rigid-rotor
expressions~\cite{15Schwenke.diatom}:
\begin{eqnarray}
 \langle  J,\Omega|\hat{J}_{z}|J,\Omega\rangle &=&\Omega, \\
 \langle  J,\Omega|\hat{J}^{2}|J,\Omega\rangle &=&J(J+1), \\
 \langle  J,\Omega\mp 1|\hat{J}_{\pm}|J,\Omega\rangle &=&\sqrt{J(J+1)-\Omega(\Omega\mp 1)} ,
\end{eqnarray}
while matrix elements of the spin operators between electronic wave functions
(omitting the `state' label for simplicity) are given by
\begin{eqnarray}
 \langle  \Lambda, S, \Sigma |\hat{S}_{z}| \Lambda, S, \Sigma  \rangle &=&\Sigma, \\
 \langle  \Lambda, S, \Sigma |\hat{S}^{2}| \Lambda, S, \Sigma \rangle &=&S(S+1), \\
 \langle  \Lambda, S, \Sigma\pm  1 |\hat{S}_{\pm}|\Lambda, S, \Sigma\rangle &=&\sqrt{S(S+1)-\Sigma(\Sigma\pm 1)}. \\
\end{eqnarray}
The coupling rules for the Hamiltonian (\ref{e:Hr}) are as follows; the first
line in Eq.~(\ref{e:Hr}) is the diagonal part of the rotational Hamiltonian,
i.e. is non-zero only for $\Delta S = \Delta \Sigma = \Delta \Lambda =0$. The
term containing $\hat{J}_\pm \hat{S}_\mp$ is called S-uncoupling and is
non-zero for $\Delta S =0 ; \Delta \Sigma = \pm 1; \Delta \Lambda =0$. The
term containing $\hat{J}_\pm \hat{L}_\mp$ is called L-uncoupling and is
non-zero for $\Delta S =0 ; \Delta \Sigma = 0; \Delta \Lambda =\pm 1$.
Finally, the term containing $\hat{S}_\pm \hat{L}_\mp$ is called
spin-electronic and is non-zero for $\Delta S =0 ; \Delta \Sigma = \pm 1;
\Delta \Lambda =\mp 1$.

%where $L$ is not really a "good" quantum number for diatomics. The relations
%(14) and (15) are correct only in the framework of Van Vleck's hypothesis of
%"pure precession"~\cite{Lefebvre-Brion-Field.book} which is valid, for
%instance, for Rydberg's states.
%The electronic, $\Lambda$-related integrals  appear in the solution only
%implicitly as matrix elements of the $\hat{L}_\pm$  and  $\hat{L}^2$
%operators in $H_{\rm r}$ as well as of the spin-orbit contributions.
Matrix elements of the orbital AM operators
$\hat{L}^{2}_{x}+\hat{L}^{2}_{y}\equiv \hat{L}^{2}-\hat{L}^{2}_{z}$ and
$\hat{L}_{\pm}$ when averaged over the electronic wave functions give rise to
$r$-dependent curves; these can be computed by \emph{ab initio} methods \cite{Colbourn1979} or
estimated semi-empirically, for example using quantum defect theory \cite{97StPuCh.diatom,01StChxx.diatom}.

The expectation value of the sum of the vibrational and the mass-polarisation
Hamiltonian using the electronic wave functions gives rise \cite{Herman1966,Kutzelnigg1997}
to the so-called Born-Oppenheimer diagonal correction (also called adiabatic
correction), which can be added to the Born-Oppenheimer PEC
$V_\mathrm{state}(r)$ if desired.

At this stage \Duo\ builds the full Hamiltonian matrix in the basis of
Eq.~(\ref{e:basis}) and using the Hamiltonian operator (\ref{e:H-tot}),
possibly complemented by supplementary terms such as spin-orbit coupling (see
section \ref{sec:additionalH} for a list of possible additional terms to the
Hamiltonian). The vibrational matrix elements
\begin{equation}
\langle {\rm state}_\lambda, \upsilon_\lambda | \hat{F}(r) | {\rm state}_\mu, \upsilon_\mu \rangle
\end{equation}
for all operators $\hat{F}(r)$ including couplings, dipole moments,
corrections etc. between different vibrational basis set functions are
computed and stored; note that in the equation above ${\rm state}_\lambda$
and ${\rm state}_\mu$ indicate different electronic states if $\lambda \neq \mu$.

At this point a basis set transformation is carried out, from the basis given by
Eq.~(\ref{e:basis}) to a symmetrized one in which the basis functions have
well-defined parity; parity (even or odd) is defined with respect to
inversion of all laboratory-fixed coordinates
\cite{93Kato.methods,Barr1975,Roeggen1971,Pack1968} and is equivalent to the
reflection operation through the molecule-fixed $xz$ plane, $
\hat{\sigma}_{v}(xz)$. The parity properties of the basis functions of
Eq.~(\ref{e:basis}) are given by Kato \cite{93Kato.methods}
\begin{equation}\label{e:parity}
 \hat{\sigma}_{v}(xz)  | {\rm state}, J, \Omega, \Lambda, S, \Sigma, \upsilon \rangle =  (-1)^{s-\Lambda+S-\Sigma +J-\Omega} | {\rm state}, J, -\Omega, -\Lambda, S, -\Sigma, \upsilon \rangle,
\end{equation}
where  $s=1$ for $|\Sigma^{-} \rangle$ states and $s=0$ for all other states.
The symmetrized basis functions are symmetric ($+$) or antisymmetric ($-$)
with respect to $ \hat{\sigma}_{v}(xz)$.
% The symmetrization is more complex than a+/-tau b
% and are given by   % will be referred to as $\phi_{\Lambda, S, \Sigma, \upsilon}^{J,\pm}$.
%\begin{equation}\begin{split}\label{e:basis:symm}
% |{\rm state}, J, |\Omega|, |\Lambda|, S, |\Sigma|, \upsilon, \tau \rangle  =  & ( | {\rm state}, \Lambda, S, \Sigma  \rangle |{\rm state},\upsilon \rangle | J,\Omega,M \rangle\\
%& +\tau | {\rm state}, \Lambda, S, \Sigma  \rangle |{\rm state},\upsilon \rangle | J,\Omega,M \rangle )/\sqrt{2}
%\end{split}\end{equation}
%where $\tau = \pm 1$ is a new parity quantum number.
%The symmetrized functions corresponding to $\tau = +1$ above are
%eigenfunctions of $\hat{\sigma}_{v}(xz)$ with eigenvalue $(-1)^{s+J+S+2\Sigma}$
%while the ones corresponding to $\tau = -1$ have eigenvalue $(-1)^{s+J+S+2\Sigma+1}$.
Use of the symmetrized basis set leads to two separate Hamiltonian
blocks with defined parities.
% For homonuclear molecules the $u/g$-symmetry
% should be preserved as well.

The two parity blocks  are then diagonalized (see Section
\ref{sec:technical} for technical details), to obtain the final rovibronic
eigenvalues $E_{\lambda}^{J,\tau}$ and corresponding eigenfunctions
$\phi_{\lambda}^{J,\tau}$, where $\tau = \pm 1$ is the parity quantum
number, $\lambda=1, 2, \ldots$ is a simple counting index.  The corresponding
rovibronic wave function $\phi_{\lambda}^{J,\tau}$ can be written as an
expansion in the basis set (\ref{e:basis})
\begin{equation}\label{basis.exp}
\phi_{\lambda}^{J,\tau} = \sum_n C_{\lambda,n}^{J,\tau} | n \rangle,
\end{equation}
where the $C_{\lambda,n}^{J,\tau}$ are expansion coefficients and $n$ here is a
shorthand index for the basis set labels `state', $J$, $\Omega$, $\Lambda$, $S$, $\Sigma$,
and $\upsilon$:
\begin{equation}\label{e:shot-hand}
| n \rangle  =  |{\rm state}, J, \Omega, \Lambda, S, \Sigma, \upsilon \rangle.
\end{equation}

As the notation above indicates, in the general case the only good quantum
numbers (i.e. labels associated with the eigenvalues of symmetry operators)
are the total angular momentum value $J$ and the parity $\tau$. Nevertheless,
\Duo\ analyzes the eigenvectors and assigns energy levels with the
approximate quantum numbers `state', $\upsilon, \Lambda$, $\Sigma$, and
$\Omega$ on the basis of the largest coefficient in the basis set expansion
(\ref{basis.exp}). It should be noted that the absolute signs of $\Lambda$
and $\Sigma$ are not well  defined, only their relative signs are. This is
related to the symmetry properties of the eigenfunctions of the
Hamiltonian~(\ref{e:H-tot}), which  are 50/50 symmetric and antisymmetric
mixtures of the $|\Lambda, \Sigma\rangle $ and $|-\Lambda,-\Sigma\rangle $
contributions. Therefore the absolute value of the quantum number $\Omega$ is
required additionally in order to fully describe the
spin-electronic-rotational contribution. In situations where some couplings
are absent some  approximate quantum numbers can become exact; for example,
in the absence of spin-orbit and spin-spin interactions the basis
functions (\ref{e:basis}) with different values of spin $S$ do not
interact and, hence, $S$ becomes a ``good'' quantum number. As another
example, without the presence of $^3 \Sigma^-$ or $^1 \Pi$ states there is no
mechanism for the ro-vibrational functions of a $^1\Sigma^+$ state to
interact with other electronic states
% I DID NOT UNDERSTAND THIS: (a part from other $^1\Sigma^+$ states through non-adiabatic interaction)
and therefore the corresponding
eigenfunctions will have well defined values of $S=\Sigma=\Lambda =0$.
% The case of the first three electronic states,
% X~$^2\Sigma^+$, A~$^2\Pi$, and B~$^2\Sigma^+$, of the AlO
% molecule~\cite{jt589}.

Table~\ref{duo:energies} gives an example of a \duo\ output with the energy
term values computed for the case of the first three electronic states, $X\
{}^{2}\Sigma^{+}$, $A\  {}^{2}\Pi$, and $B\  {}^{2}\Sigma^{+}$, of AlO
\cite{jt589}.

\begin{table}
\begin{center}
\caption{
Sample \Duo\ energy output for AlO~\cite{jt589}. The energy is given in
cm$^{-1}$, and the exact (\texttt{J}, \texttt{n}, \texttt{parity}) and
approximate (\texttt{state}, \texttt{v} ($\upsilon$), \texttt{lambda}
($\Lambda$), \texttt{spin} ($S$), \texttt{sigma} ($\Sigma$), and
\texttt{omega} ($\Omega$)) quantum numbers. The final  column contains labels
of the electronic states as given by the user and the separator \texttt{||}
is to facilitate selecting the energy entries in the program
output.\label{duo:energies}}
\hrule
\begin{myverbatim}
       J      N        Energy/cm  State   v  lambda spin   sigma   omega  parity
       0.5    1          0.000000   1     0   0     0.5     0.5     0.5   +    ||X2SIGMA+
       0.5    2        965.435497   1     1   0     0.5     0.5     0.5   +    ||X2SIGMA+
       0.5    3       1916.845371   1     2   0     0.5     0.5     0.5   +    ||X2SIGMA+
       0.5    4       2854.206196   1     3   0     0.5     0.5     0.5   +    ||X2SIGMA+
       0.5    5       3777.503929   1     4   0     0.5     0.5     0.5   +    ||X2SIGMA+
       0.5    6       4686.660386   1     5   0     0.5     0.5     0.5   +    ||X2SIGMA+
       0.5    7       5346.116382   2     0   1     0.5    -0.5     0.5   +    ||A2PI
       0.5    8       5581.906844   1     6   0     0.5     0.5     0.5   +    ||X2SIGMA+
       0.5    9       6066.934830   2     1   1     0.5    -0.5     0.5   +    ||A2PI
       0.5   10       6463.039443   1     7   0     0.5     0.5     0.5   +    ||X2SIGMA+
       0.5   11       6778.997803   2     2   1     0.5    -0.5     0.5   +    ||A2PI
       0.5   12       7329.427637   1     8   0     0.5     0.5     0.5   +    ||X2SIGMA+
       0.5   13       7483.145675   2     3   1     0.5    -0.5     0.5   +    ||A2PI
       0.5   14       8159.170405   2     4   1     0.5    -0.5     0.5   +    ||A2PI
       0.5   15       8201.467744   1     9   0     0.5     0.5     0.5   +    ||X2SIGMA+
       0.5   16       8857.266385   2     5   1     0.5    -0.5     0.5   +    ||A2PI
\end{myverbatim}
\hrule
\end{center}
\end{table}

\subsection{Solution of the uncoupled vibrational problem}\label{solution.SE}
The main method of solving the radial equation used by \Duo\ is the
so-called sinc DVR (discrete variable representation);
this method (or closely related ones) has been independently applied to the
one-dimensional Schr{\"o}dinger equation by various authors
\cite{Lund1984,Guardiola1982,Guardiola1982a,Colbert1992}. %but was made popular by Colbert and Miller \cite{Colbert1992}.

\begin{center}
\begin{figure}
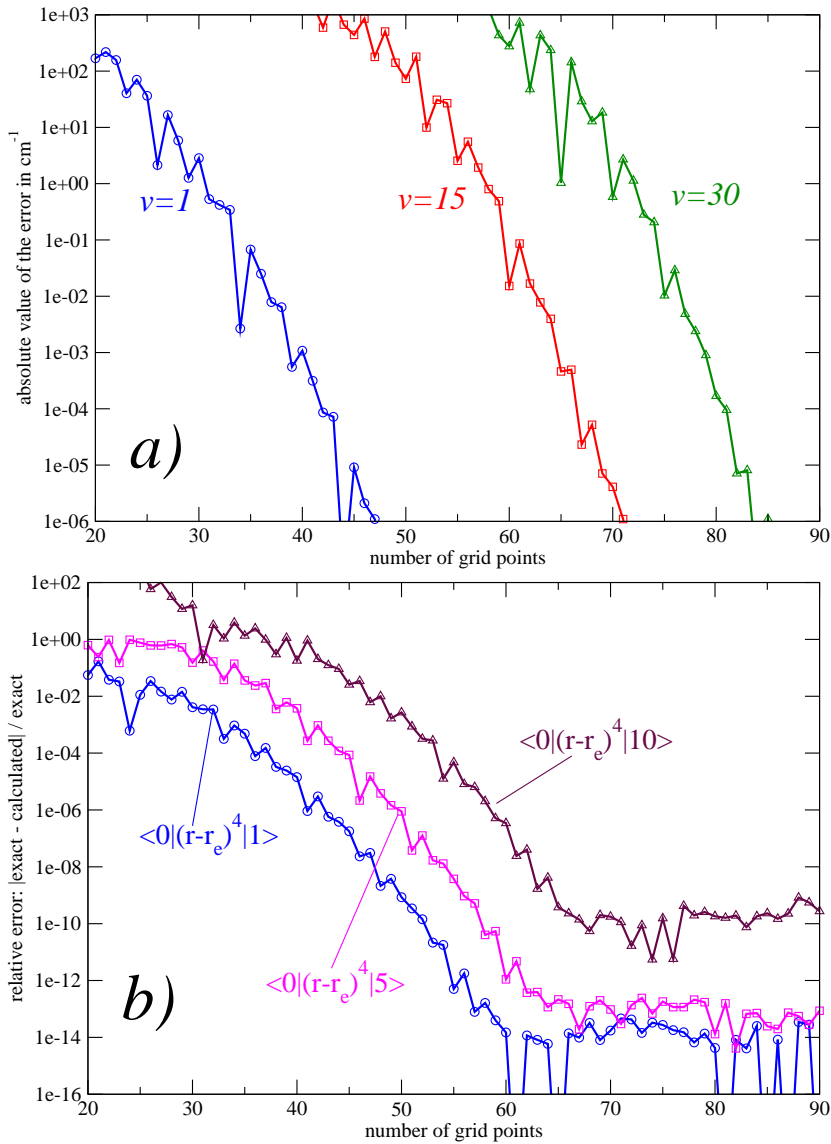

\includegraphics[angle=0, width=0.80\textwidth]{./J0.morse.eps}
\includegraphics[angle=0, width=0.80\textwidth]{./J0.morse.ME.eps}
\caption{Illustrative examples of the fast convergence for energies (plot \emph{a}) and matrix elements (plot \emph{b}) using the sinc DVR method; in both cases the rate of convergence is approximately exponential with respect to the number of grid points. Results are for a Morse potential which approximately models the ground electronic state of the CO molecule, $r_e = 1.1283$~\AA, $D_e = 90674$~\cm, $\omega = 2169.814$~\cm\ with atomic masses for carbon and oxygen. A uniformly spaced grid was used, keeping fixed $r_\mathrm{min}=0.7 $~\AA, $r_\mathrm{max}=2.0 $~\AA. In plot \emph{a)} we show absolute errors for the $v=1$, $v=15$ and $v=30$ energy levels; in plot \emph{b)} we show relative errors of matrix elements of the type $\langle 0 | (r-r_e)^4| v\rangle$; the flattening of the error for large numbers of grid points is due to the numerical error present in floating point calculations (see text). \label{fig.J0.morse}}
\end{figure}
\end{center}

In this method the $r$ coordinate is truncated to an interval
$[r_\mathrm{min}, r_\mathrm{max}]$ and discretized in a grid of $N_{p}$
uniformly spaced points $r_i = r_\mathrm{min} + i \Delta r$  (where $i\in
[0,N_p-1]$) with grid step $\Delta r  = (r_\mathrm{max} -
r_\mathrm{min})/(N_\mathrm{p}-1)$. The Schr{\"o}dinger Eq.~(\ref{1d-SE})
is then transformed to an ordinary matrix eigenvalue problem
\begin{equation} \label{e:Hvib.matrix}
  (\mathbf{T} + \mathbf{V} ) \vec{\psi}_\upsilon = E_\upsilon \vec{\psi}_\upsilon,
\end{equation}
where $\mathbf{T}$ is the matrix representing the kinetic energy and is given in the sinc method by \cite{Colbert1992,Tannor.book}
\begin{equation}\label{T.matrix.sinc.infty}
T_{ij} = \frac{\hbar^2}{2 \mu (\Delta r)^2} \left\{ \begin{array}{ll}
  \frac{\pi^2}{3} & i=j\\
  2\frac{(-1)^{i-j}}{(i-j)^2} & i\neq j\\
\end{array} \right.
\end{equation}
and $\mathbf{V} = \mathrm{diag}[V(r_0), V(r_1), \ldots, V(r_{N_p-1})]$ while
the vector $\vec{\psi}_\upsilon$ contains the values of $\psi(r)$ at the grid
points. The resulting $N_p \times N_p$ real symmetric matrix $\mathbf{H}$ is
then diagonalized (see section \ref{sec:technical} for details).
The sinc DVR method %ideally possesses
usually provides very fast  (faster than polynomial) convergence
of the calculated energies and wave functions with respect to the number of the grid points, $N_{p}$.
% As a guideline a grid step $\Delta r = 0.01$~\AA should be sufficient to converge to better
% than 10$^{-4}$~\cm\ all vibrational levels up to $v=80$.
Figure~\ref{fig.J0.morse} \emph{a)} shows the convergence for three $J=0$ energy levels of a Morse
potential, showing a rate of convergence approximately exponential with respect to the number of
grid points.
% Note, however, this
% requires the potential to be continuous until the $N_{p}$-th derivative with
% respect to $r$.
\Duo\ obtains  all integrals over vibrational coordinates by summation over
the grid points:
\begin{equation}\label{rect.rule}
  \int_{r_\mathrm{min}}^{r_\mathrm{max}} \psi_\lambda(r) F(r) \psi_\mu(r) \ \mathrm{d}r = \Delta r \sum_{i=0}^{N_p-1} \psi_\lambda(r_i) F(r_i) \psi_\mu(r_i) .
\end{equation}
The rectangle rule is simple and exponentially accurate for
integration over infinite range of functions which decay fast
(exponentially or faster) and which do not have singularities in the
complex plane close to the real axis~\cite{Trefethen2014}.  We
illustrate in fig.~\ref{fig.J0.morse} \emph{b)} the quick convergence
of matrix elements of the type $\langle 0 | (r-r_e)^4| v\rangle$ for a
Morse potential; analytical formulae for matrix elements of this kind
are available from the literature \cite{Gallas1980,Rong2003} and were
used to obtain exact reference values.  In plot \ref{fig.J0.morse}
\emph{b)} it is apparent that the accuracy of matrix elements does not
improve beyond a certain value; for example, the matrix elements
$\langle 0 | (r-r_e)^4| 10\rangle$ always has less than about 10
significant digits no matter how many points are used.  This behaviour
is completely normal and expected when performing floating-point
calculations with a fixed precision; \Duo\ uses double precision
numbers with a machine epsilon $\varepsilon = 2 \times 10^{-16}$
\cite{Higham1993} and the expected relative error due to the finite
precision in the sum given by eq.~(\ref{rect.rule}) is given by,
indicating with $S$ the value of the sum performed with infinite
accuracy and with $\hat{S}$ the value obtained with finite accuracy:
\begin{equation}\label{error.sum}
 \frac{|S-\hat{S}|}{S}  \approx \varepsilon N_p \frac{ \sum_{i}| \psi_\lambda(r_i) F(r_i) \psi_\mu(r_i)|}{| (\sum_{i}\psi_\lambda(r_i) F(r_i) \psi_\mu(r_i) )|}
\end{equation}
The expression above implies that whenever the matrix element of a function $F$ comes out
very small with respect to the value of $|F|$ significant digits will be lost;
there are techniques such as Kahan compensated summation \cite{Higham1993} which
reduce the error above by a factor $N_p$ but these have not been implemented
at this time.

A prime example of this situation if given by the line intensities of
very high vibrational overtones; in a recent study
\citet{Medvedev2015} %should decrease % approximately exponentially.
observed that matrix elements of the type $|\langle 0 | \mu(r)| v
\rangle|$ for the CO molecule when computed with double precision
floating-point arithmetic decrease approximately exponentially (as
expected on the basis of theoretical models and as confirmed by
quadruple precision calculations) for $v \lesssim 25 $, when they
reach the value of about $10^{-15}$~D.  This situation is fully
expected on the basis of the considerations above but it should never
constitute a problem in practice. %; although
% very high overtones are overestimated, they are
% still assigned a very small value (they are 10^{30} times smaller than strong lines)
% so it should not matter at all.

Apart from the sinc DVR, \Duo\ implements finite-difference (FD) methods for
solving the uncoupled vibrational problem, where the kinetic energy operator
$\mathbf{T}$ in Eq.~(\ref{e:Hvib.matrix}) can be approximated using, for
example, a 5-point central FD5 formulae:
\begin{equation}\label{T.matrix.5point}
T_{ij} = \frac{\hbar^2}{2 \mu (\Delta r)^2} \left\{ \begin{array}{ll}
  5/2  & i=j,\\
 -4/3  & |i- j| = 1,\\
  1/12 & |i- j| = 2.\\
\end{array} \right.
\end{equation}
and furthermore with $T_{1\, 1} = T_{N_p N_p}=29/12$. Note that the expression above gives incorrect results for the first two and last two grid points,
but this does not matter as long as the grid truncation parameters $r_\mathrm{min}$ and $r_\mathrm{max}$
are chosen so that $\psi_\upsilon \approx 0$ near the borders of the grid.

The formulae (\ref{T.matrix.5point}) lead to a symmetric pentadiagonal
banded matrix, which can in principle be diagonalized more efficiently than a
generic dense matrix. However, the convergence of the eigenvalues
$E_\upsilon$ is much slower, with error decreasing as $(\Delta r)^4$ instead
of $e^{-\alpha/(\Delta r)}$.

% To improve the
% convergence of both eigenvalues and matrix elements derived by the FD5 method
% the two points Richardson's $h^4$-extrapolation to zero step~\cite{NR2007},
% $h\to 0$, is exploited. The energy error can be estimated by the
% integral~\cite{CCerrorFD, Meshkov2008}:
% \begin{eqnarray} \label{FD5}
% \Delta E_\upsilon ~=~\frac{\hslash^2} {2\mu}\left( \frac{h^4} {90}\right) \int_{r_\mathrm{min}}^{r_\mathrm{max}}
% \left[\frac{d[(V_{\rm state}(r)-E_\upsilon)\psi_\upsilon(r)]}{dr}\right]^2 dr.
% \end{eqnarray}

\subsection{Levels lying close to dissociation}\label{sec:non-uniform}
\begin{center}
\begin{figure}
\includegraphics[angle=0, width=0.95\textwidth]{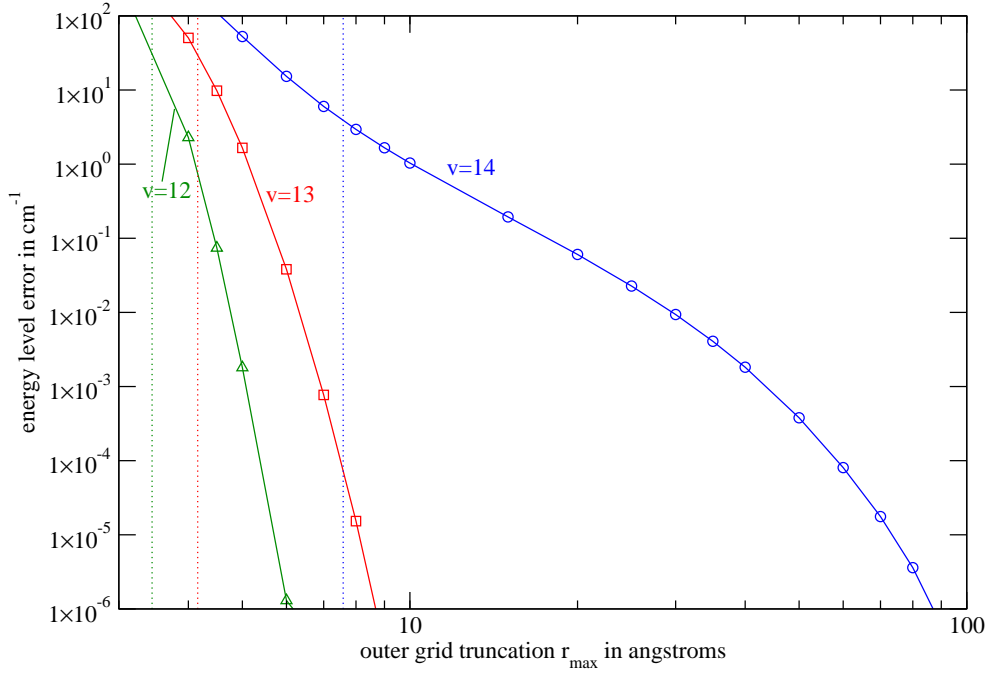}
\caption{Illustrative example of the effect of the outer grid truncation parameter $r_\mathrm{max}$ on energy levels close to dissociation. Data are relative to a Morse potential with $D_e = 12728.237432$~\cm, $\omega_e=1751.124844$~\cm, $r_e = 1$~\AA, $\mu = 1$~Da. This potential supports 15 bound states ($v=0$ to $v=14$) and we consider in this example the three highest-energy ones, with energies $E_{12} = -250.7130$~\cm, $E_{13} = -65.2068$~\cm, $E_{14} = -0.1000$~\cm. In all calculation we fixed $r_\mathrm{min}=0.1$~\AA\ and the grid step $\Delta r$ to 0.05~\AA. The dotted vertical lines are the outer turning points for the three states, i.e. the points $r_\mathrm{out}$ such that $V(r_\mathrm{out})=E_v$; the error in the computed energy levels is expected to decrease exponentially when $r_\mathrm{max} > r_\mathrm{out}$. The plot shows that to converge the last energy level $E_{14}$ a very large $r_\mathrm{max}$ is required, which in turn leads to a large number of grid points when they are uniformly spaced. Specifically, to converge $E_{12}$ to $10^{-6}$~\cm\ it is sufficient to choose $r_\mathrm{max} > 6$~\AA, leading to 120 points; for $E_{13}$ we need $r_\mathrm{max} > 9$~\AA\ and 180 points; for $E_{14}$ we need $r_\mathrm{max} > 90$~\AA\ and 1500 points. \label{fig.rmax}}
\end{figure}
\end{center}

A general requirement for convergence %of the both sinc DVR and FD methods
is that both the inner and the outer grid truncation values $r_\mathrm{min}$ and $r_\mathrm{max}$
should be chosen such that $V_{\rm state}(r_\mathrm{min})$ and $V_{\rm state}(r_\mathrm{max}) $
are both much larger than $E_\upsilon$.
A problem arises when one is trying to converge states very close to
the dissociation limit, as such loosely bound states can
extend to very large values of $r$ and
therefore require an excessive number of points when a uniformly spaced grid
is used; this is illustrated in fig.~\ref{fig.rmax}.
Excited states of alkali diatoms such as Li$_2$ \cite{Coxon2006}, Na$_2$ \cite{Qi2007}
or K$_2$ \cite{Falke2006}
constitute an important class of systems for which large $r_\mathrm{max}$ are needed;
such systems are prime choices for studies of ultracold atoms and molecules
\cite{Tiemann2013} and often require grids extending
up to several tens \cite{Coxon2006,Qi2007} or even
hundreds \cite{Falke2006} of Angstroms.

In such cases it may be beneficial to use a non uniform grid; %For this
\Duo\ implements the adaptive analytical mapping approach of Meshkov
\emph{et al} \cite{Meshkov2008} and offers several mapping choices, which are
described in the manual. However, at this time support for non uniform grids
should be considered experimental and they cannot be used in combination
with the sinc DVR method but only with the less efficient
5-point finite-difference one. Indicatively we recommend considering
non-uniform grids only when $r_\mathrm{max}$ is required to be
larger than $\approx$ 50~\AA.

\subsection{States beyond the dissociation limit}\label{sec:quasibound}
\begin{center}
\begin{figure}
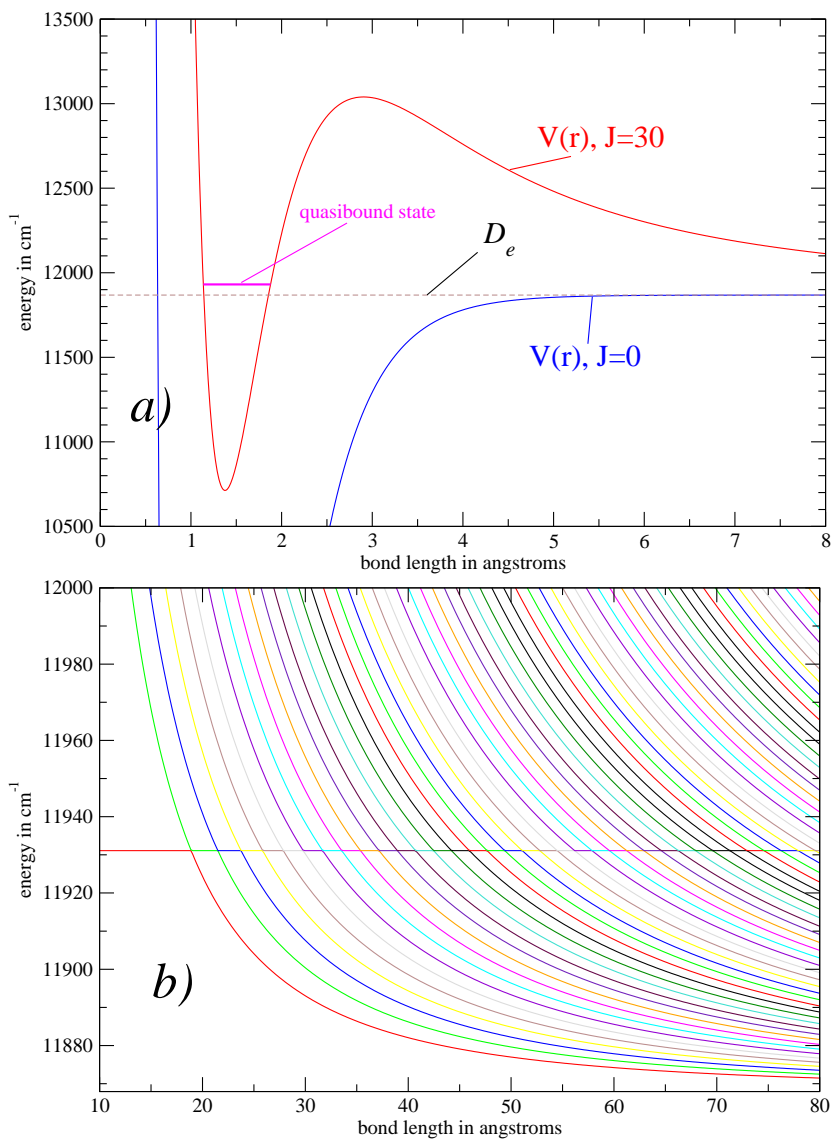

\includegraphics[angle=0, width=0.80\textwidth]{./plot_morse_J30.eps}
\includegraphics[angle=0, width=0.80\textwidth]{./plot_resonance_5_to_100_dense.eps}
\caption{Example of a quasibound, orbiting resonance state. Plot \emph{a)}: Morse potential for $J=0$ and $J=30$ (same parameters as in fig.~\ref{fig.rmax}); the potential for $J=30$ has a local maximum higher than the dissociation limit $D_e$ and supports one quasibound state with energy $E=$11~931.1~\cm. Plot \emph{b)}: eigenvalues for $E > D_e$ as a function of the outer grid truncation $r_\mathrm{max}$. The quasibound state manifests itself as a series of avoided crossings. \label{fig:Morse.J30}}
\end{figure}
\end{center}

Potential curves with local maxima higher than the dissociation limit of the potential for $r \to +\infty$
may support shape resonances, i.e. metastable states in which the two atoms are trapped
for a finite time in the potential well but eventually dissociate.
Such states are also known as quasibound or tunnelling predissociation states.
For $J>0$ the rotational potential will practically always introduce such a maximum,
and the corresponding quasibound levels are known as orbiting resonances or rotationally
predissociating states, see fig.~\ref{fig:Morse.J30} \emph{a)} for an example.
Several techniques have been developed to deal with quasibound states, most notably
in the context of diatomic molecules
by Le~Roy and co-workers \cite{LeRoy1971,LeRoy1978,Connor1981,Pryce1994,Riss1995,Cizek1996,Sidky1999,Huang2003,Huang2003a}.
% , i.e. to find their energies and lifetimes.
At the moment \Duo\ does not provide any explicit functionality
to treat quasibound states,
% to single them out from the continuum of states beyond dissociation.
although we plan to rectify this deficiency in future versions.

Nevertheless, long-lived quasibound states (i.e., narrow resonance)
can be identified using the present version of \Duo\ by using the so-called stabilization
method \cite{Hazi1970,Simons1981,Levebvre1985,GarciaSucre1986,Mandelshtam1993,Martin1999}.
In one version of
this approach energy levels are computed for increasing values of the outer
grid truncation $r_\mathrm{max}$ and then plotted as function of $r_\mathrm{max}$;
quasibound states manifests themselves by being relatively stable with respect
to increase of $r_\mathrm{max}$  and undergo a series of avoided crossings,
see fig.~\ref{fig:Morse.J30} \emph{b)} for an example. From an analysis of these curves
it is also possible to compute the lifetime of the quasibound state \cite{Mandelshtam1993}.

% Alternatively one can keep $r_\mathrm{max}$ fixed to an appropriately
% large value, perform calculation with increasing numbers of grid points $N_p$
% and plot the energy levels as functions of $N_p$.
% Also in this case quasibound states reveal themselves
% by being relatively stable with respect to increase of $N_p$ and undergo
% avoided crossings.

\subsection{Printing the wave functions}\label{sec:printing.WFs}
Both the $J=0$ vibrational basis functions $| \mathrm{state}, v
\rangle$, see eq.~(\ref{e:basis}), and the final ($J >0 $,
electronically coupled or both) rovibronic wave functions coefficients
$C_{\lambda,n}^{J,\tau}$, see eq.~(\ref{basis.exp}), can be written to a
file for further analysis, e.g. for plotting purposes or for the
computation of $g$ factors; see the manual for details.

\subsection{Convergence of rotationally excited states}\label{sec:convergence.J}
In our approach $J > 0$ calculations are performed using a basis expansion in terms
of the $J=0$ wave functions. %and we comment in this section on the convergence of this expansion.
As a guideline it was found by numerical experimentation
that in order to obtain converged results for rotationally excited states
up to $\upsilon_\mathrm{max}$ %and $J_\mathrm{max}$
one has to use a vibrational $J=0$ basis set of size only
slightly larger than $\upsilon_\mathrm{max}$ and that a reasonable minimum value for
the size of the vibrational basis set is given by $1.25 \times \upsilon_\mathrm{max} +2$.
For example, %in a single-surface calculation
to converge rotationally excited levels up to $v=30$ it should be sufficient to
use a vibrational basis set of size 40.

\subsection{Additional terms in the Hamiltonian}\label{sec:additionalH}
\Duo\ supports the inclusion of a number of terms  additional to the
non-relativistic Hamiltonian (\ref{e:H-tot}) caused by spin-orbit
$\hat{H}_{\rm SO}$, spin-rotational $\hat{H}_{\rm SR}$, spin-spin
$\hat{H}_{\rm SS}$ and $\Lambda$-doubling $\hat{H}_{\rm LD}$
interactions~\cite{Lefebvre-Brion-Field.book,93Kato.methods,87BrChMe.methods,88DaAbSa.C2,79BrMexx.methods,Richards.book}:

\begin{enumerate}
\item The Breit-Pauli spin-orbit operator $\hat{H}_{\rm SO}$
    \cite{Marian2001,Fedorov2003,Veseth1970,Richards.book} has non-zero matrix elements
    between electronic states obeying the following coupling rules
    \cite{Lefebvre-Brion-Field.book}:
      $\Delta S=0,\pm 1$; $\Delta \Lambda=0, \pm 1$; $\Delta \Omega =0$;   %, , $\Delta \Sigma=0, \pm 1$.
      if $\Delta S =0$ and $\Sigma' = \Sigma'' =0$ the matrix elements is
      zero (this last rule implies that singlet-to-singlet matrix elements
      are zero); $\Sigma^+$ electronic states may have non-zero matrix
      elements with $\Sigma^-$ states but $\Sigma^\pm \leftrightarrow
      \Sigma^\pm$ matrix elements are zero; finally, in case of homonuclear
      diatomics, only $g \leftrightarrow g$ and $u \leftrightarrow u$
      matrix elements are non-zero.

 The diagonal SO matrix elements $ \langle \Lambda, S, \Sigma |\hat{H}_{\rm
 SO}| \Lambda, S, \Sigma  \rangle $ determine the spin-orbit splitting of a
 multiplet $^{2S+1}\Lambda$, where $S>0$ and $\Lambda>0$. Both diagonal and
 off-diagonal matrix elements of the spin-orbit Hamiltonian can be obtained
 as functions of $r$ using quantum chemistry
 programs. %Alternatively, for heavy atoms the corresponding
% relativistic effective core potentials (ECP) including the SO splitting
% part could be employed.

\item The nonzero diagonal and off-diagonal matrix elements of $\hat{H}_{\rm SR}$ operator are given by
\begin{eqnarray}\label{e:SR:diag}
\langle \Lambda, S, \Sigma |\hat{H}_{\rm SR}| \Lambda, S, \Sigma  \rangle &=& \frac{\hbar^2}{2 \mu r^2}\gamma^{\rm SR}(r)\left[\Sigma^2 -S(S+1) \right], \\
\langle \Lambda, S, \Sigma |\hat{H}_{\rm SR}| \Lambda, S, \Sigma \pm 1  \rangle &=& \frac{\hbar^2}{4 \mu r^2}\gamma^{\rm SR}(r)
\left[S(S+1) -\Sigma (\Sigma \pm 1)\right]^{1/2}\nonumber \\
&\times & \left[J(J+1) -\Omega (\Omega \pm 1)\right]^{1/2}
\end{eqnarray}
where $\gamma^{\rm SR}(r)$ is a dimensionless function of $r$.

\item The diagonal matrix elements of the $\hat{H}_{\rm SS}$ operator are
    taken in the phenomenological form
\begin{eqnarray}\label{e:SS}
\langle \Lambda, S, \Sigma |\hat{H}_{\rm SS}| \Lambda, S, \Sigma \rangle &=&
\frac{2}{3} \lambda^{\rm SS}(r)\left[3\Sigma^2 - S(S+1)\right]
\end{eqnarray}
Both  $\gamma^{\rm SR}(r)$ and $\lambda^{\rm SS}(r)$ functions can be obtained either \ai\ or empirically.

\item The lambda-doubling (LD) couplings for a $^{2S+1}\Pi$ state in the
    $\Lambda$-representation (Eq.~(\ref{e:basis})) are of the following
    three types \cite{79BrMexx.methods}:
\begin{equation} \label{e:Hlambda-doubling}
\hat{H}_{\rm LD} = \frac{1}{2} \alpha_{opq}^{\ LD }(r)  \left( \hat{S}_+^2+\hat{S}_{-}^{2}  \right)
- \frac{1}{2} \alpha_{p2q}^{\rm LD }(r)  \left( \hat{J}_{+}\hat{S}_{+} + \hat{J}_{-}\hat{S}_{-} \right)
+\frac{1}{2}  q^{\rm LD }(r)  \left( \hat{J}_{+}^2+\hat{J}_{-}^2  \right),
\end{equation}
where $\alpha_{opq}^{\rm LD }$ and $\alpha_{p2q}^{\rm LD }$ are related to
the conventional terms as given by \citet{79BrMexx.methods}:
\begin{eqnarray}
\nonumber
  \alpha_{opq}^{\rm LD } &=&  o^{\rm LD }+p^{\rm LD }+q^{\rm LD } \\
  \alpha_{p2q}^{\rm LD } &=&  p^{\rm LD }+2q^{\rm LD }.
\end{eqnarray}

% To account for the Born-Oppenheimer breakdown (BOB) effect (regular
% $L$-uncoupling) caused by the remote states~\cite{lr07}, the rotational
% kinetic energy operator ${\hbar^2}/{2 \mu r^2}$ in Eq.~\eqref{e:Hr} can be
% replaced by
% \begin{equation}
% \label{e:dist}
% \frac{\hbar^2}{2 \mu r^2}\left [ 1 + g^{\rm BOB}(r)\right ],
% \end{equation}
% where the unitless BOB functions $g^{\rm BOB}(r)$ can be represented as an
% empirically defined expansion.
\item It is now well-established that, at least for $^1\Sigma$ states, the
    small shifts to energy levels due to non-adiabatic interactions with
    remote states (as opposed to near-degenerate ones) can be accurately
    modelled by modifying the vibrational and rotational energy operators
    in the Hamiltonian
    \cite{Pachucki2008,Watson2004,LeRoy1999,Herman1998,Bunker1977,Herman1966};
    specifically, the vibrational energy operator in Eq.~(\ref{e:H-vib}) is
    replaced by
\begin{equation}\label{BOB:vib}
-\frac{\hbar^2}{2 \mu} \frac{\mathrm{d}}{\mathrm{d} r} \left( 1 + \beta(r) \right) \frac{\mathrm{d}}{\mathrm{d} r}
\end{equation}
while the rotational kinetic energy operator ${\hbar^2}/{2 \mu r^2}$ in
Eq.~\eqref{e:Hr} should be replaced by
\begin{equation}\label{e:dist}
\frac{\hbar^2}{2 \mu r^2}\left( 1 + \alpha(r)\right).
\end{equation}
The functions $\alpha(r)$ and $\beta(r)$
are sometimes referred to as Born-Oppenheimer breakdown (BOB) curves \cite{02LeHuxx.diatom} and
can also be interpreted as introducing position-dependent vibrational
and rotational masses; they are sometimes expressed in terms of
the dimensionless $g$-factor functions $g_\mathrm{v}$ and $g_\mathrm{r}$ by
$\alpha(r) = (m_e / m_p) g_\mathrm{r}(r)$ and $\beta(r) = (m_e / m_p) g_\mathrm{v}(r)$.
The rotational $g_\mathrm{r}$ function can be determined experimentally by analysis
of the Zeeman splitting of energy levels due to an external magnetic field \cite{Ogilvie2000}.
\end{enumerate}

\subsection{Representation of the couplings}\label{sec:mat.el.extra}
\Duo\ assumes that the coupling matrix elements and the transition dipole
moments are given in the representation of the basis
functions~(\ref{e:basis}) corresponding to Hund's case (a). In this
representation the $\hat{L}_z$ component is diagonal and has a signed value
$\Lambda$ (see Eq.(13)) and therefore it will be referred to as the
$\Lambda$-representation. It can be shown that by choosing appropriate phase
factors for the electronic wave functions $| {\rm state}, \Lambda, S, \Sigma
\rangle$ all coupling matrix elements in the $\Lambda$-representation can be
made real; note that in this representation the electronic wave functions are
complex numbers, as they contain a factor of the kind $e^{i \Lambda \phi}$,
where $\phi$ is the angle corresponding to rotation around the $z$ axis
\cite{Barr1975}. On the other hand quantum chemistry programs such as
{\sc Molpro}~\cite{MOLPRO2012} normally work with real wave functions
and consequently compute matrix elements in this representation, which we
call Cartesian as the electronic wave functions are
ultimately expressed in terms of atom-centred Cartesian components
\cite{Marian2001} $|x\rangle$, $|y\rangle$, $|z\rangle$, $|xy\rangle$
\emph{etc}.

\Duo\ can accept input in either the Cartesian or the $\Lambda$-representation.
For the Cartesian-representation \Duo\ will then transform these inputs to
the $\Lambda$-representation as follows:
\begin{equation}\label{e:lambda-transf}
 |\pm\Lambda\rangle = \left[ C_{1}^{\pm |\Lambda|} |1\rangle + C_{2}^{\pm |\Lambda|} |2\rangle \right],
\end{equation}
where $1$ and $2$ denote Cartesian components $x, y, z, xy, \ldots$ that correspond to the $A_1, B_1$ and $A_2, B_2$
Abelian point group symmetries, respectively. $C_{\alpha}^{\Lambda}$ are the elements of the unitary transformation from the Cartesian to the $\Lambda$-representation. The obvious way to reconstruct this transformation is to diagonalize the Cartesian representation of the $\hat{L}_z$ matrix. Thus the transformed matrix elements in the
$\Lambda$-representation are given by
\begin{equation}\label{e:cart-to-Lambda}
 \langle \Lambda \Sigma | \hat{A} | \Lambda\p \Sigma\p \rangle =
 \sum_{\alpha,\beta=0,1,2} (C_{\alpha}^{\Lambda})^{*}  C_{\beta}^{\Lambda\p}
 \langle \alpha \Sigma | \hat{A} | \beta \Sigma\p \rangle
\end{equation}
or, in tensorial form $\hat{\boldmath A}^{\rm Duo} = \hat{\boldmath C}^{\dagger} \hat{\boldmath A} \hat{\boldmath C}$ where $\alpha=\beta=0$ correspond to a $^{2S+1}\Sigma$ ($\Lambda=0$) electronic state with $C_{0}^{0} = 1$.

In principle all Cartesian matrix elements $\langle \alpha \: \Sigma |
\hat{A} | \beta \: \Sigma\p \rangle$ must be provided to perform the
transformation in Eq.~(\ref{e:cart-to-Lambda}). However, by means of the
coupling rules all non-zero matrix elements $\langle \alpha \Sigma | \hat{A}
| \beta \Sigma\p \rangle$ can be related to only one, non-zero reference
matrix element. For example, the matrix element $\langle \Lambda  =
\Sigma = 0 |H^{\rm SO}| \Lambda\p = \Sigma \p = \pm 1 \rangle$ between
$^{1}\Sigma$ and $^{3}\Pi$ is zero because it corresponds to a simultaneous
change of $\Lambda$ and $\Sigma$ by $\pm 1$. This property together with the
help of Eq.~\eqref{e:lambda-transf} allows one to use the non-zero spin-orbit
matrix elements $\langle 0, \Sigma  = 0 |H^{\rm SO}| 2, \Sigma  = 1 \rangle $
as a reference and to reconstruct all other non-zero Cartesian component by
\begin{eqnarray}
 \langle 0, \Sigma  = 0 |H^{\rm SO}| 1, \Sigma  = 1 \rangle  = - \frac{C_{2}}{C_{1}} \langle 0, \Sigma  = 0 |H^{\rm SO}| 2, \Sigma  = 1 \rangle ,
\end{eqnarray}
as required for Eq.~\eqref{e:cart-to-Lambda}.

Off-diagonal matrix elements of the various operators included
into our model, i.e. the various couplings between electronic states,
are subject in actual calculations
to arbitrary changes of sign due to the sign indeterminacy of the
electronic wave functions computed at different geometries. Often the phases of each \ai\ coupling
$F(r)$ have to be post-processed in order to provide a consistent, smooth
function of $r$. It is important that the relative phases between different
elements preserved. This issue is illustrated graphically by~\citet{jt589},
where different \ai\ coupling curves of AlO obtained with {\sc Molpro}
were presented. Transition dipole moment functions, discussed in the next
section, also may exhibit phase changes ~\cite{jt573}, which %These  %and other artifacts
should be corrected using the same phase convention used for other %coupling
matrix elements \citep{jt589}.

\subsection{Computational considerations}\label{sec:technical}
\Duo\ uses %makes use, when necessary, of
the matrix diagonalization routines DSYEV or,
optionally, DSYEVR from the LAPACK library \citep{99AnBaBi.method}.
The subroutine DSYEVR uses the multiple
relatively robust representations algorithm and is expected to be faster than
DSYEV, which is based on the QR algorithm~\cite{Demmel2008, VanZee2014};
however, the current version of DSYEVR is poorly parallelized and therefore
not recommended for parallel environments.

The dimension of the final rovibrational Hamiltonian matrix depends on
the number of vibrational functions selected, the number of electronic
states present, the spin multiplicities of the electronic states and
the $J$ quantum number. For example, for $N$ electronic states,
$N_\upsilon$ vibrational functions are retained for each of them and
denoting with $\overline{m}$ the average spin multiplicity, the size
of the Hamiltonian matrix is approximately given by $N \times
N_\upsilon \times \overline{m}$ and the size of the $\tau=\pm$ parity
matrix to be diagonalized is half of this value.  The size of each
block of the Hamiltonian reaches dimensions of the order of a
thousand only for rather complicated cases (e.g., $N=10$,
$N_\upsilon=40$ and $\overline{m}=5$) and consequently the time taken
to compute the energy levels for a given $J$ is usually only a small
fraction of a second.

\section{Line intensities and line lists}\label{sec.intensities}
The Einstein coefficient $A_{fi}$ (in $1/s$) for a transition $\lambda_f \gets \lambda_i$ is computed as
\begin{eqnarray}
A_{fi} &=& \frac{64\times 10^{-36} \pi^4}{3 h} \, (2 J_i+1) \, \tilde{\nu}^3 \sum_{t=-1,0,1} \left|  \sum_{n_i,n_f}
\left(C_{\lambda_f,n_f}^{J_f,\tau_f}\right)^{*} C_{\lambda_i,n_i}^{J_i,\tau_i} \right. \\
&& \left.  (-1)^{\Omega_i}
\left(\begin{array}{ccc}
J_i & 1 & J_f\\
\Omega_i & t & -\Omega_f
\end{array}
\right) \langle \upsilon_f|  \bar\mu_{t}^{f,i}(r) |  \upsilon_i \rangle
\right|^2 ,
\end{eqnarray}
where $\bar{\mu}_t$ $(t=-1,0,1)$ are the electronically averaged body-fixed components of the electric dipole moment (in Debye) in the irreducible representation
\begin{eqnarray}
\bar{\mu}_0 = \bar{\mu}_z;\qquad
\bar{\mu}_{\pm 1} &=&  \mp \frac{1}{\sqrt{2}} (\bar{\mu}_{x} \pm i \bar{\mu}_{y}),
\end{eqnarray}
and the index $n$ is defined by Eq.~(\ref{e:shot-hand}). The vibrationally
averaged transition dipole moments $\langle \upsilon_f|
\bar{\mu}_{t}^{f,i}(r) | \upsilon_i \rangle$ are computed using the
vibrational wave functions $|\upsilon\rangle \equiv \psi_{\upsilon}(r)$ .

The absorption line intensity is then given by %\red{Comment from Andrey: check this formula please}
\begin{equation}
\label{e:int}
 I(f\gets i) = \frac{g_{\rm ns} (2 J_f+1) A_{fi}}{8 \pi c \tilde{\nu}^2}  \frac{e^{-c_2 \tilde{E}_i/T } \left( 1-e^{-c_2\tilde{\nu}_{if}/T} \right)}{Q(T)},
\end{equation}
where $Q(T)$ is the partition function defined as
\begin{equation}
\label{e:pf}
  Q(T) = g_{\rm ns} \sum_{i} (2J_{i}+1) e^{-c_2\tilde{E}_i/T},
\end{equation}
$g_{\rm ns}$ is the nuclear statistical weight factor, $c_2= hc / k_B$
is the second radiation constant, $\tilde{E}_i = E_i/h c $ is the term value, and $T$ is
the temperature in K. For heteronuclear molecules $g_{\rm ns}$ is a total
number of combinations of nuclear spins as given by $g_{\rm ns} = (2 I_{b}+1)
(2I_{a}+1),$ where $I_a$ and $I_b$ are the corresponding  nuclear spins. For
a homonuclear molecules, these combinations are distributed among the four
symmetries $+s$, $-s$, $+a$, $-a$, where $+/-$ is the parity of the molecule
with respect to $\sigma_v$ and $s/a$ is the  property of the total rovibronic
wave function to be symmetric/asymmetric upon upon
inversion~\cite{93Kato.methods}. In the representation of $C_{2v}$ point
group symmetry, this corresponds to $A_1$, $A_2$, $B_1$, and $B_2$. Thus, for
the case $I\equiv I_a=I_b$ two different values $g_{\rm ns}$ are
necessary and these depend on whether the nuclei are fermions ($I$ is half-integer)
or bosons ($I$ integer)~\cite{Herzberg.book,06SiJaRo.method}:
\begin{eqnarray}
g_{\rm ns} = \left\{
\begin{array}{ll}
\frac{1}{2} \left[ (2 I+1)^2- (2 I+1)  \right], & {\rm Fermi},s \; {\rm and} \;{\rm Bose},a\\
\frac{1}{2} \left[ (2 I+1)^2+ (2 I+1)  \right], & {\rm Fermi},a \; {\rm and} \; {\rm Bose},s.
\end{array}
\right.
\end{eqnarray}
For example, carbon $^{12}$C has $I=0$ and therefore
for the C$_2$ molecule $g_{\rm ns}$ are 1 for $A_1$, $A_2$ and 0 for $B_1$, $B_2$ states,
respectively.

The computed Einstein $A$ coefficients can be used to compute
radiative lifetimes of individual states and cooling functions in a
straightforward manner \cite{jt624}.

% The selection rules are $$  + \leftrightarrow - $$ $$ \Delta J = 0, \pm 1, \ \ \ {\rm with}  0 \nleftrightarrow 0 $$
%for heteronuclear and $$A_1 (+s) \leftrightarrow A_2 (-s) \;\; , B_1 (+a) \leftrightarrow B_2 (-a) $$
\subsection{Line list format}
A line list is defined as a catalogue of transition frequencies and
intensities \cite{jt548}. In the basic ExoMol format \cite{jt548},
adopted by \duo, a line list consists of two files: `States' and
`Transitions'; an example for the molecule AlO is given in
Tables~\ref{t:levels} and \ref{t:trans}. The `States' file contains
energy term values supplemented by the running number $n$, total
degeneracy $g_n$, rotational quantum number $J_n$ (all obligatory
fields) as well as quantum numbers $\upsilon$, $\Lambda$, parity
($\pm$), $\Sigma$, $\Omega$ and the electronic state label (e.g.
\verb!X2Sigma+!). The `Transitions' file contains three obligatory
columns, the upper and lower state indexes $n_f$ and $n_i$ which are
running numbers from the `State' file, and the Einstein coefficient
$A_{fi}$. For the convenience we also provide the wavenumbers
$\tilde{\nu}_{if}$ as the column 4. The line list in the ExoMol format
can be used to simulate absorption or emission spectra for any
temperature in a general way. Note that ExoMol format has recently
been significantly extended \cite{jtexo} but structure of the
States and Transitions file has been retained.

\begin{table}
\caption{Extract from the output `State' file produced by \Duo\ for the $^{27}$Al$^{16}$O molecule~\cite{jt598}.}
\label{t:levels}
\begin{center}
\footnotesize
\tabcolsep=5pt
\begin{tabular}{rrrrrrlrrrr c rrr}
\hline
     $n$ & \multicolumn{1}{c}{$\tilde{E}$} &  $g$    & $J$  &  \multicolumn{1}{c}{$+/-$} &  \multicolumn{1}{c}{$e/f$} & State & $\upsilon$    & $|\Lambda|$& $|\Sigma|$ & $|\Omega|$& \\
\hline
           1  &       0.000000   &      12   &      0.5  &   +  &   e   &  \verb!X2SIGMA+ !&     0   &   0   &      0.5    &     0.5  \\
           2  &     965.435497   &      12   &      0.5  &   +  &   e   &  \verb!X2SIGMA+ !&     1   &   0   &      0.5    &     0.5  \\
           3  &    1916.845371   &      12   &      0.5  &   +  &   e   &  \verb!X2SIGMA+ !&     2   &   0   &      0.5    &     0.5  \\
           4  &    2854.206196   &      12   &      0.5  &   +  &   e   &  \verb!X2SIGMA+ !&     3   &   0   &      0.5    &     0.5  \\
           5  &    3777.503929   &      12   &      0.5  &   +  &   e   &  \verb!X2SIGMA+ !&     4   &   0   &      0.5    &     0.5  \\
           6  &    4686.660386   &      12   &      0.5  &   +  &   e   &  \verb!X2SIGMA+ !&     5   &   0   &      0.5    &     0.5  \\
           7  &    5346.116382   &      12   &      0.5  &   +  &   e   &  \verb!A2PI     !&     0   &   1   &      0.5    &     0.5  \\
           8  &    5581.906844   &      12   &      0.5  &   +  &   e   &  \verb!X2SIGMA+ !&     6   &   0   &      0.5    &     0.5  \\
           9  &    6066.934830   &      12   &      0.5  &   +  &   e   &  \verb!A2PI     !&     1   &   1   &      0.5    &     0.5  \\
          10  &    6463.039443   &      12   &      0.5  &   +  &   e   &  \verb!X2SIGMA+ !&     7   &   0   &      0.5    &     0.5  \\
          11  &    6778.997803   &      12   &      0.5  &   +  &   e   &  \verb!A2PI     !&     2   &   1   &      0.5    &     0.5  \\
          12  &    7329.427637   &      12   &      0.5  &   +  &   e   &  \verb!X2SIGMA+ !&     8   &   0   &      0.5    &     0.5  \\
          13  &    7483.145675   &      12   &      0.5  &   +  &   e   &  \verb!A2PI     !&     3   &   1   &      0.5    &     0.5  \\
          14  &    8159.170405   &      12   &      0.5  &   +  &   e   &  \verb!A2PI     !&     4   &   1   &      0.5    &     0.5  \\
          15  &    8201.467744   &      12   &      0.5  &   +  &   e   &  \verb!X2SIGMA+ !&     9   &   0   &      0.5    &     0.5  \\
          16  &    8857.266385   &      12   &      0.5  &   +  &   e   &  \verb!A2PI     !&     5   &   1   &      0.5    &     0.5  \\
          17  &    9029.150380   &      12   &      0.5  &   +  &   e   &  \verb!X2SIGMA+ !&    10   &   0   &      0.5    &     0.5  \\
          18  &    9535.195842   &      12   &      0.5  &   +  &   e   &  \verb!A2PI     !&     6   &   1   &      0.5    &     0.5  \\
          19  &    9854.882567   &      12   &      0.5  &   +  &   e   &  \verb!X2SIGMA+ !&    11   &   0   &      0.5    &     0.5  \\
          20  &   10204.019475   &      12   &      0.5  &   +  &   e   &  \verb!A2PI     !&     7   &   1   &      0.5    &     0.5  \\
          21  &   10667.668381   &      12   &      0.5  &   +  &   e   &  \verb!X2SIGMA+ !&    12   &   0   &      0.5    &     0.5  \\
          22  &   10864.560220   &      12   &      0.5  &   +  &   e   &  \verb!A2PI     !&     8   &   1   &      0.5    &     0.5  \\
          23  &   11464.897083   &      12   &      0.5  &   +  &   e   &  \verb!X2SIGMA+ !&    13   &   0   &      0.5    &     0.5  \\
          24  &   11519.212123   &      12   &      0.5  &   +  &   e   &  \verb!A2PI     !&     9   &   1   &      0.5    &     0.5  \\
          25  &   12156.974798   &      12   &      0.5  &   +  &   e   &  \verb!A2PI     !&    10   &   1   &      0.5    &     0.5  \\
          26  &   12257.694655   &      12   &      0.5  &   +  &   e   &  \verb!X2SIGMA+ !&    14   &   0   &      0.5    &     0.5  \\
          27  &   12793.671660   &      12   &      0.5  &   +  &   e   &  \verb!A2PI     !&    11   &   1   &      0.5    &     0.5  \\
          28  &   13030.412255   &      12   &      0.5  &   +  &   e   &  \verb!X2SIGMA+ !&    15   &   0   &      0.5    &     0.5  \\
          29  &   13421.583651   &      12   &      0.5  &   +  &   e   &  \verb!A2PI     !&    12   &   1   &      0.5    &     0.5  \\
          30  &   13790.933964   &      12   &      0.5  &   +  &   e   &  \verb!X2SIGMA+ !&    16   &   0   &      0.5    &     0.5  \\
\hline
\end{tabular}
\end{center}

\mbox{}\\
{\flushleft
$n$:   State counting number.     \\
$\tilde{E}$: State energy in \cm. \\
$g$: State degeneracy.            \\
$J$:   Total angular momentum. \\
$+/-$:   Total parity. \\
$e/f$:   Rotationless parity. \\
State:   Electronic state label. \\
$\upsilon$:   State vibrational quantum number. \\
$\Lambda$:   Absolute value of $\Lambda$ (projection of the electronic angular momentum). \\
$\Sigma$:   Absolute value of $\Sigma$ (projection of the electronic spin). \\
$\Omega$:   Absolute value of $\Omega=\Lambda+\Sigma$ (projection of the
total angular momentum).}
\end{table}

\begin{table}
\caption{Sample extracts from the output `Transition' files produced by \duo\ for the $^{27}$Al$^{16}$O molecule~\cite{jt598}.}
\label{t:trans}
\begin{tabular}{rrrr}
\hline\hline
\multicolumn{1}{c}{$n_f$}	&  \multicolumn{1}{c}{$n_i$} 		& 		\multicolumn{1}{c}{$A_{fi}$}   &  \multicolumn{1}{c}{$\tilde{\nu}_{fi}$} \\
\hline\hline
         173     &      1 &\texttt{ 4.2062E-06      }&$      1.274220   $   \\
         174     &      1 &\texttt{ 1.3462E-02      }&$    966.699387   $   \\
         175     &      1 &\texttt{ 1.3856E-02      }&$   1918.099262   $   \\
         176     &      1 &\texttt{ 9.7421E-03      }&$   2855.450672   $   \\
         177     &      1 &\texttt{ 1.2511E-06      }&$   3778.740157   $   \\
         178     &      1 &\texttt{ 1.1544E-02      }&$   4687.891466   $   \\
         179     &      1 &\texttt{ 6.7561E+02      }&$   5346.110326   $   \\
         180     &      1 &\texttt{ 4.1345E+00      }&$   5583.130067   $   \\
         181     &      1 &\texttt{ 2.4676E+03      }&$   6066.924557   $   \\
         182     &      1 &\texttt{ 3.5289E+01      }&$   6464.257469   $   \\
         183     &      1 &\texttt{ 4.6960E+03      }&$   6778.981670   $   \\
         184     &      1 &\texttt{ 1.9771E+02      }&$   7330.641321   $   \\
         185     &      1 &\texttt{ 6.1540E+03      }&$   7483.122722   $   \\
         186     &      1 &\texttt{ 4.8062E+03      }&$   8159.737396   $   \\
         187     &      1 &\texttt{ 1.9401E+03      }&$   8202.080179   $   \\
\hline
\end{tabular}
\noindent \mbox{}\\ $n_f$: Upper state counting number.\\ $n_i$: Lower state counting number.\\
$A_{fi}$:  Einstein-A coefficient in s$^{-1}$.\\ $\tilde{\nu}_{fi}$:
Transition wavenumber in \cm\ (optional).

\end{table}

\section{Inverse problem}\label{s:refinement}
The inverse problem consists in finding the potential energy and coupling
curves which best reproduce a given set of energy levels,  $E_{i}^{\rm
(obs)}$, or frequencies (i.e., differences between energy levels), typically
extracted from experiment. In the following we will call this
optimization process \emph{empirical refinement}. %LL:actually  in the following there is no reference to 'empirical refinement'

\subsection{Implementation}

% Empirical refinement is possible when curves are given with any of the analytical function forms listed in Section~\ref{s:functions}; curves
% available only by spline interpolation cannot be empirically refined and are not considered  below.

The refinement problem can be formulated as a non-linear least-squares
problem where one seeks to minimize the objective function
\cite{96DeScxx.methods}:
%(see \red{CHECK
%Dennis, J.E. and Schnabel, R.B. (1983). Numerical Methods for Unconstrained
%Optimization and Nonlinear Equations. Prentice-Hall, Eaglewood Cliffs.}):
\begin{equation}
F  = \sum_i \left[E_{i}^{\rm (obs)}-E_{i}^{\rm (calc)}(a_n)\right]^2 w_{i},
\end{equation}
where $E_{i}^{\rm (calc)}(a_n)$ are the calculated energies or frequencies
and implicitly depend on the parameters $a_1,a_2,\ldots $ defining the
potential and coupling curves. The $w_i > 0$ are weighting factor assigned to
each value and may be chosen as $1/\sigma_i^2$ where $\sigma_i$ is the
experimental uncertainty of $E_{i}^{\rm (obs)}$. The input weights are
automatically renormalized by \Duo\ so that $\sum_{i} w_i = 1$.

\Duo\ uses the non-linear conjugate gradient method for the optimization; in
particular, the linearized least-square problem is solved by default using
the LAPACK subroutine DGELSS, although the alternative built-in subroutine
LINUR is also available. For each curve appearing in \Duo\ it is possible to
specify if any given parameter should be refined (fitted) or should be kept
fixed to the value given in the input file.
The first derivatives with respect to the fitting parameters $a_n$ required
for the non-linear least squares are computed using
finite differences with a step size $\Delta a_n$ taken as 0.1\%\ of the
initial values $a_n$ or 0.001 if  $a_n$ is initially zero.
% The starting values are chosen as the \ai\ values when available.

\subsection{Constrained minimization}
In order to avoid unphysical behaviour and also to avoid problems when
% underdetermined problems where
the amount of experimental data provided is insufficient for determining all
the parameters, the shapes of the curves can be contrained to be as close as
possible to some reference curves provided in the input (typically \ai\ ones)
~\cite{79TiArxx.method, 03YuCaJe.PH3, jt503, 14SzCsxx.methods}. This is done
by including into the fitting objective function not only differences of the
computed energy levels but also differences between the refined curves
$V^{\lambda, \rm (calc)}$ and the reference ones $V^{\lambda, \rm (ref)}$
% the corresponding reference curve, $V^{\lambda}(r)$,
% as grid points $V^{\lambda}_i$,
as follows:
\begin{equation}
\label{e:F-all}
F = \sum_i (E_{i}^{\rm (obs)}-E_{i}^{\rm (calc)})^2 w_{i}^{\rm en} +
    \sum_{\lambda} d_{\lambda}  \sum_k (V_{k}^{\lambda, \rm (ref)}-V_{k}^{\lambda, \rm (calc)})^2 w_{k}^{\lambda},
\end{equation}
where $\lambda$ refers to the $\lambda$-th curve, $k$ counts over the grid points,
$w_{i}^{\lambda}$ are the corresponding weight factors of the individual points normalized to one %within each property,
and $d_{\lambda}$ are further weight factors defining the relative importance
of the corresponding curve. The weights in Eq.~\eqref{e:F-all} are normalized
as follows:
\begin{equation}
  \sum_{i} w_i^{\rm en} +  \sum_{\lambda}   \sum_k d_{\lambda} w_{k}^{\lambda}  = 1.
\end{equation}

When minimizing the functional given by Eq.~\eqref{e:F-all} it is important to
control the correctness of the match %correlation
between the experimental and theoretical levels as they appear in the
corresponding observed (`obs.') and calculated  (`calc.') lists. It is
typical in complex fits involving close-lying electronic states that
the order of the computed energy levels in a ($J$, $\tau$) block
%/frequences %LL I suggest we drop in the discussion reference to frequences, to make the text more readable.
changes during the empirical refinement.
In order to follow these changes and update the
positions of the experimental values in the fitting set, we use the quantum
numbers to identify the corresponding quantities by locking to their initial
values. Since the experimentally assigned quantum numbers may not
agree with the ones used by \Duo\ --- which are based on Hund's case $a)$ ---
% do not always agree with the
% theoretical ones, for this purposes
each experimental energy level (or frequency)
$E_{i}^{\rm (obs)}$ is automatically labelled
by \Duo\ with the following %theoretical
quantum numbers: $J$, parity $\tau$($\pm$), `state', $\upsilon$, $|\Lambda|$,
$|\Sigma|$ and  $|\Omega|$;
this set of six quantum numbers %represents a
%signature for an experimental entry $E_{i}^{\rm (obs)}$ and
is then used for
matching with a calculated counterpart $E_{i}^{\rm (calc)}$. %from all the
%computed energies (or frequencies).
Note that only the absolute values of $\Lambda$, $\Sigma$ and $\Omega$ are
used for this purpose, as their sign is undefined in the general case.
%To control the assignment within a specific energy/wavenumber region an energy
%threshold $E^{\rm thresh}$ is used. The initial assignment can be
%reconstructed using energy closeness criteria.

\subsubsection{Morphing}
% If the curves to be refined have a complicated but physically reasonable
% shape, it can be `morphed' \cite{99MeHuxx.methods, 99SkPeBo.methods, jt589}
The curves to be refined can also be `morphed' \cite{99MeHuxx.methods, 99SkPeBo.methods, jt589}
by scaling them by a function $H(r)$, so that the refined function $F(r)$
at a given grid point $r_i$ is given by
\begin{equation}
  F(r_i)=  H(r_i) F^{\rm initial}(r_i),
\label{eq:morph}
\end{equation}
where $F^{\rm ai}(r_i)$ is initial function specified in the input file (e.g., obtained
by \ai\ methods). %obtained from a sparser \ai\ grid of geometries using a splines interpolation.
For this kind of empirical refinement the curves $F^{\rm ai}$ do not
necessarily have to be specified by a parametrised analytical form but can
also be provided %by may be specified
as a spline interpolant as described in
Section \ref{sec:splines}.

The morphing function $H(r)$ is typically represented by a simple polynomial,
see \citet{jt589} for an example.
The morphing approach is an alternative way of constraining the refined
properties to the reference curve %, usually \ai\ curves and thus useful when the
and is especially useful when
experimental information is sparse. %The morphing procedure is of particular
% use for coupling functions when a simple analytical forms is not
% appropriate.

\section{Types of functional forms}\label{s:functions}

\subsection{Analytical representations}
A number of functional forms  are currently
available in \Duo\ to specify %any property as an
$r$-dependent curves (e.g., potential energy curves,
%non-adiabatic couplings, BOB corrections and %permanent and transition
dipole moment curves) as parametrised analytical functions.
In the following $T_{\rm e}$ will represent
the value of the potential at the equilibrium geometry $r_{\rm e}$.

\begin{enumerate}
\item Expansion in Dunham variables \cite{Dunham1932}:
\begin{equation}
\label{e:dunham}
V(r) = T_{\rm e}+a_0  \, y_{\rm D}^2(r) \left( 1+   \sum_{i\ge 1} a_i y_{\rm D}^i(r) \right),
\end{equation}
where
$$
y_{\rm D}(r) = \frac{r-r_{\rm e}}{r_{\rm e}}.
$$

\item Taylor polynomial expansion:
\begin{equation}
V(r) = T_{\rm e} + \sum_{i\ge 0} a_i \, (r-r_{\rm e})^i.
\end{equation}

\item Simons-Parr-Finlan (SPF) \cite{73SiPaFi.method,Fougere1966}
    expansion:
\begin{equation}
\label{e:SPF}
V(r) = T_{\rm e}+a_0 y_{\rm SPF}^2(r) \left( 1+   \sum_{i\ge 1} a_i y_{\rm SPF}^i(r) \right),
\end{equation}
where
$$
y_{\rm SPF}(r) = \frac{r-r_{\rm e}}{r}.
$$

\item Murrell-Sorbie (MS) \cite{75SoMuxx.method}:
$$
V(r) = T_{\rm e} + (A_{\rm e}-T_{\rm e}) e^{-a (r-r_{\rm e})} \left( 1+ \sum_{i\ge 1} a_i (r-r_{\rm
e})^i \right),
$$
where $A_{\rm e}$ is the asymptote of $V(r)$ at $r \to +\infty$ relative to

$T_{\rm e}$ of the lowest electronic state, related to the commonly
used dissociation energy of the given electronic state $D_{\rm e}$  =
$A_{\rm e}-T_{\rm e}$.

\item Chebyshev polynomial expansion \cite{11BuKlNi.KCs, 13KnRuTi.Mg2}:
\begin{equation}
 \label{e:Chebyshev}
   V(r) = [T_{\rm e} + D_{\rm e}] -\frac{\sum_{i=0} a_i T_k(y_p)}{1\,+\,\left(r/r_{\rm ref}\right)^n},
 \end{equation}
in which $\,n\,$ is a positive integer and $\,T_k(y)\,$ are the Chebyshev
polynomials of the first kind defined in terms of the reduced variable
$\,y_p(r)\in [-1,1]\,$:
\begin{eqnarray}
y_p(r;r_{\rm min},r_{\rm ref})=\frac{r^p-r_{\rm ref}^p}{r^p+r_{\rm ref}^p-2r_{\rm min}^p}
\end{eqnarray}
with $p$ as the fixed parameter. This form guarantees the correct
long-range (LR) behaviour at $r\to \infty$:
\begin{equation}\label{eq:LR}
V(r)\to u_{\rm LR}(r) = \sum_{n} \frac{C_n}{r^n}
\end{equation}
where the $C_n$ are the long-range coefficients.

\item Perturbed Morse Oscillator (PMO)~\cite{Huffaker1976,Huffaker1976a, Dwivedi1977,Huffaker1978,Huffaker1980,Huffaker1981}:
\begin{equation}\label{e:Morse}
    V(r)=T_{\rm e}+ (A_{\rm e}-T_{\rm e})  y_{\rm M}^2 +  \sum_{i=1} a_i  y_{\rm M}^{i+2},
\end{equation}
where
$$
 y_{\rm M} = 1-\exp\left(-\beta (r-r_{\rm e}\right).
$$
When $a_i=0$ the form reduces to the Morse potential, otherwise the
asymptotic value of the potential is $A_{\rm e} + \sum_i a_i$.

%This functional form should work fairly well to describe regions where the potential is
%mildly anharmonic, but it cannot describe well the long range van der Waals region.

\item Extended Morse Oscillator (EMO)~\cite{EMO,LeRoy2006}:
\begin{equation}\label{eq:EMO}
V(r)=T_{\rm e} + (A_{\rm e}-T_{\rm e})\left( 1~-~\exp\left\{-\beta_{\rm EMO}(r) (r-r_{\rm e})\right\} \right)^2,
\end{equation}
which has the form of a Morse potential with a exponential tail and a
distance-dependent exponent coefficient
\begin{equation}
\beta_{\rm EMO}(r)~=~ \sum_{i=0} a_i y_p^{\rm eq}(r)^i,
\end{equation}
expressed as a power series in the reduced
variable~\cite{84SuRaBo.method}:
\begin{eqnarray}\label{eq:ypEQ}
y_p^{\textrm{eq}}(r)~=~\frac{r^p-r_{\rm e}^p}{r^p+r_{\rm e}^p}.
\end{eqnarray}

\item Morse Long-Range (MLR) function~\citep{LeRoy2006,LeRoy2007,lr07,11LeHaTa.method}:
\begin{equation}\label{eq:MLR}
V(r) = T_{\rm e}+ (A_{\rm e}-T_{\rm e}) \left(1 - \frac{u_{\textrm{LR}}(r)}
{u_{\textrm{LR}}(r_e)}~ \exp\!\left\{ -\beta_{\rm MLR}(r) y_p^{\textrm{eq}}(r)\right\}\right)^2,
\end{equation}
where the radial variable $y_p^{\rm eq}$ in the exponent is given by
Eq.~(\ref{eq:ypEQ}), the long-range potential $u_{\textrm{LR}}(r)$ by
Eq.~(\ref{eq:LR}) while the exponent coefficient function
\begin{eqnarray}
\beta_{\rm MLR}(r) = y_p^{\rm{ref}}(r)\, \beta_{\infty}~ +~ \left[1 -y_p^{\textrm{ref}}(r)\right]~\sum_{i=0} a_i[y_q^{\textrm{ref}}(r)]^i
\end{eqnarray}
is defined in terms of two radial variables which are similar to $y_p^{\rm
eq}$, but are defined with respect to a different expansion centre
$r_\textrm{ref}$, and involve two different powers, $p$ and $q$. The above
definition of the function $\beta_{\rm MLR}(r)$ means that:
\begin{eqnarray}
\beta_{\rm MLR}(r\to\infty) ~\equiv~ \beta_{\infty}~ =~ \ln[2D_{\rm e}/u_{\textrm{LR}}(r_{\rm e})].
\end{eqnarray}

%\item XXXXX \red{CHECK THE NAME}:
%\begin{equation}
%V(r)=T_{\rm e} + (A_{\rm e}-T_{\rm e}) \frac{(1-e^{-\beta(r)})^2} {(1-e^{-\beta(+\infty)})^2},
%\end{equation}
%where
%$$
%\beta(r) = \sum_{k=0}^N c_k y(r)^k
% $$
% and
% $$
% y(r)= \frac{r-r_0}{r+r_0}.
% $$
% Note that $y(r)$  ranges from $-1$ (r=0) to $0$ ($r=+\infty$), so that $\beta(r=+\infty) = \sum_k c_k$.
%The factor in the denominator guarantees that $V(r)$ is  $D_{\rm e}$ at the
%dissociation.

% in particular for very large r this function form will go to  $ D_e + A/r
%$ where the constant  $A$ is determined by all the expansion coefficient. For neutral
%molecules the asymptotic form is  $ D_e + A/r^6 $ so the EMO form cannot describe
%dissociation well (goes to a constant too slowly).

\item \v{S}urkus-polynomial expansion~\cite{84SuRaBo.method}:
\begin{equation}
V(r) = T_{\rm e} + (1-y_p^{\textrm{eq}}) \sum_{i\ge 0} a_i [y_p^{\textrm{eq}}]^i + y_p^{\textrm{eq}} a_{\rm inf},
\end{equation}
where $y_p^{\textrm{eq}}$ is the \v{S}urkus variable (\ref{eq:ypEQ}) and
$a_{\rm inf}$ is the asymptote of the potential at $r\to \infty$.

\item \v{S}urkus-polynomial expansion with a damping
    function~\cite{84SuRaBo.method}:
\begin{equation}
V(r) =  T_{\rm e} + \left[ (1-y_p^{\textrm{eq}}) \sum_{i\ge 0} a_i [y_p^{\textrm{eq}}]^i + y_p^{\textrm{eq}} a_{\rm inf}\right] f^{\rm damp} + t^{\rm damp} (1- f^{\rm damp}),
\end{equation}
where the damping function is defined by
$$
f^{\rm damp} = 1-\tanh[\alpha(r-r_0)],
$$
and $t^{\rm damp}$, $r_0$ and $\alpha$ are parameters.

% \item Morse Lennard Jones / Morse Long range [see Level's manual, JCP 125 164310 (2006)]
% $$
% V(r) = T_{\rm e} + D_{\rm e} z^2,
% $$
% where
% $$
%  z = 1-e^{-\phi(r) (r-r_{\rm e})}\left[ 1+\epsilon_6\left(-\left(\frac{r_0}{r}\right)^6\right) \right] \left[ 1+\epsilon_8\left(-\left(\frac{r_0}{r}\right)^8\right) \right],
% $$
% where $r_0$, $\epsilon_6$, $\epsilon_8$ are expansion parameters and $\phi(r)$ is a polynomial expansion
%  in terms of the \v{S}urkus variable $y$ \red{CHECK THAT THE SAME DEFINITION OF SURKUS IS USED}
% $$
% \phi(r) = \sum_{k\ge 0} b_k y^k .
% $$

% \item An external, `user-defined' function be also easily provided via the
%     `USER' keyword.

\end{enumerate}

\subsection{Numerical representations}\label{sec:splines}
Any $r$-dependent curve $F(r)$ %used in \Duo\
% A functional form of a potential energy or any other property, here
% generically indicated as the multiplicative function $F(r)$,
can be specified as a list of data points $\{r_k, F(r_k)\}, k =1, \ldots,
N_p$ for a range of geometries. % $r_k \in [r_1,r_2]$
% which does not necessarily
% coincide with the integration interval $[r_{\rm min},r_{\rm max}]$.
\Duo\ will then automatically interpolate or extrapolate the data points whenever
necessary. The interpolation within the specified range is performed either
by using cubic or quintic splines. More specifically,
\Duo\ uses natural cubic splines in the form given in Ref.~\cite{NR2007}
or natural quintic splines based on an adaptation of the routine
QUINAT \cite{Herriot1983, Herriot1976}.
Quintic splines are used by default, as they generally provide
quicker convergence of the interpolant with respect to
the number of points given; however, they may lead to spurious oscillations
between the data points, especially for non-uniform grids.
The number of data points should be
$\geq 4$ for cubic splines and $\geq 6$ for quintic splines. It is sometimes
useful to interpolate a transformed set, such as $(r_i, r_i^2 F_i)$ or
$(1/r_i, F_i)$, see, for example, Refs. \cite{Poll1966,LeRoy1968}
or the discussion by \citet{Lodi.phd}; this
feature is not yet implemented in \Duo. %\ but will be in future versions.

%\paragraph{Extrapolation}
When necessary curves are extrapolated at short range (i.e., in the
interval $[r_\mathrm{min},r_1]$) using one of the following functional forms:
\begin{eqnarray}
  f_1(r) &=& A + B/r, \\
  f_2(r) &=& A r + B r^2,\\
  f_3(r) &=& A + B r,
\end{eqnarray}
where the constants $A$ and $B$ are found by requiring that the functions $f_i$ go through the first two data points. By default the functional form $f_1$
is used for potential energy curves, form $f_2$ is used for the transition dipole moments and form $f_3$ (linear extrapolation) for all other curves (for example spin-orbit couplings). The default choices should be appropriate in most cases.

Similarly, whenever necessary curves are extrapolated at long range (i.e., in the interval $[r_{N_k}, r_\mathrm{max}]$) by fitting the last two data points to
\begin{eqnarray}
 f_4(r)  &=& A + B/r^6\\
 f_5(r)  &=& A + B r, \\
 f_6(r)  &=& A/r^2 + B/r^3.
\end{eqnarray}
These functional forms were chosen to describe the behaviour at long ranges
of the potential energy curves [$f_4(r)$], of the curves corresponding to the
electronic angular momenta  $L_x(r)$, $L_y(r)$, $L_{\pm}(r)$ [$f_5(r)$]; for
all other cases (including transition dipole moment) the functional form
$f_6$ is assumed.
Note that form $f_4(r)$ is appropriate in many but by no means in all cases, see
ref.~\cite{LeRoy1970} for details on the the asymptotic behaviour
of the potential energy curves at large $r$.

Similarly, Ref.~\cite{Goodisman1963} discusses the asymptotic forms of molecular
(diagonal) dipole functions and states the correct limits have the form
$A + B r^m$ with $m=3$ or $5$ for $r\to 0$ and $A + B/r^m$ with $m=4$ or $7$ for
$r\to \infty$ so that our extrapolation forms do not have the correct asymptotic forms.
Nevertheless usually extrapolation is performed quite far from the aymptotic region
so that using the theoretically correct form is not required nor, indeed,
beneficial in such cases.

It should be noted that the extrapolation procedures described introduce a
small discontinuity in the first derivatives %of the corresponding property
at the switching points $r_1$ or $r_{N_p}$. In some situations these artifacts %should have little
%the convergence of the finite-difference methods used to solve the radial equations
% low-energy states lying far  but they
could become important, e.g.
for very loosely bound states such as those discussed in section \ref{sec:non-uniform}.
In such cases it is recommeded to use an analytical representation for the
potential with an appropriate long-range behaviour, e.g. the Morse long-range
form given by Eq.~(\ref{eq:MLR}).
%  However, the exponential convergence of the DVR method will not be maintained.

\section{Program inputs and structure}
\label{s:program}

\begin{center}
\begin{figure}
\epsfxsize=12.5cm \epsfbox{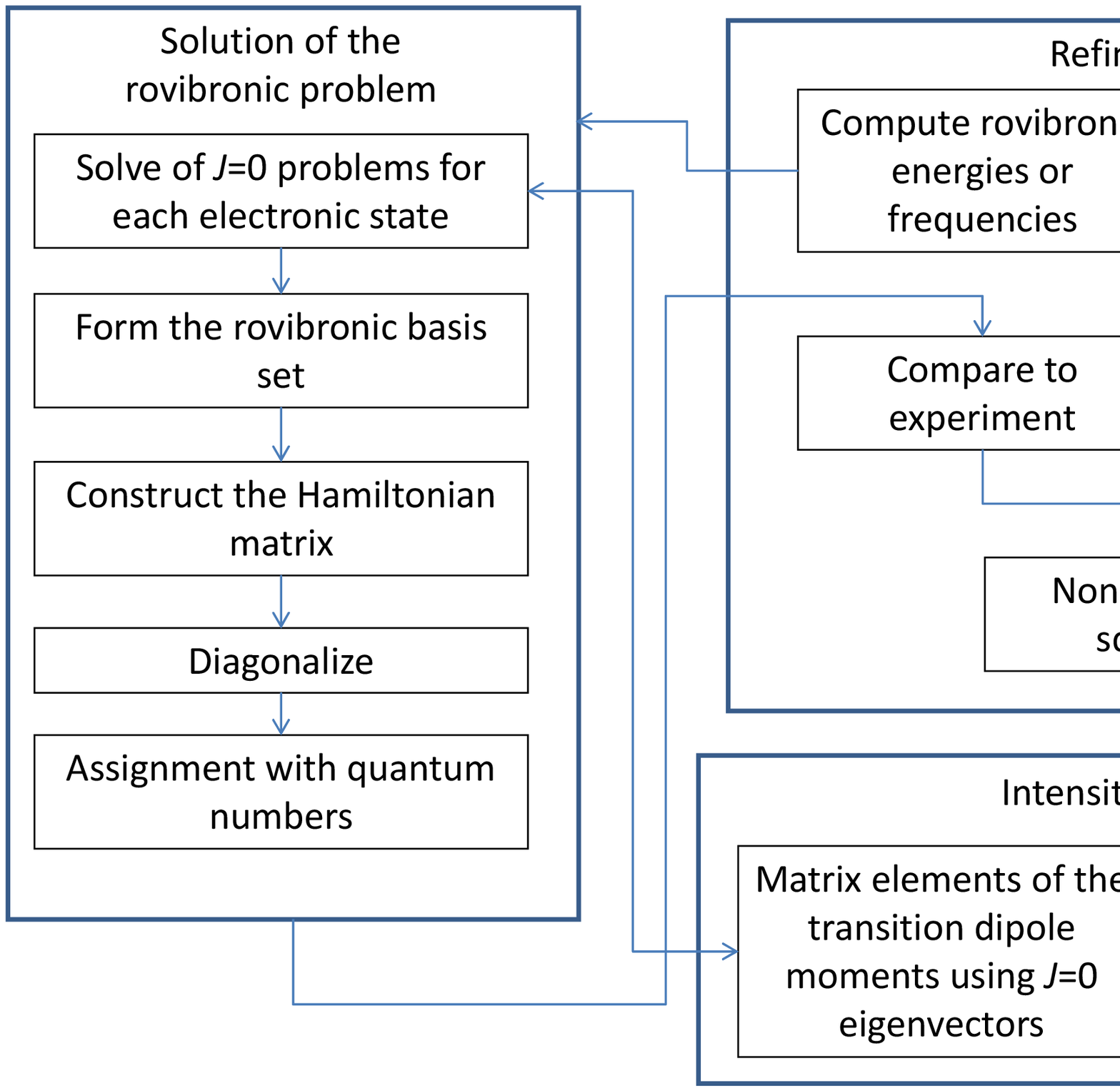}
\caption{\Duo\ program structure. \label{f:program}}
\end{figure}
\end{center}

The \Duo\ calculation set up is specified by an input file in the plain text
(ASCII) format. %, where a specific task is described as a diatomic `project'.
The input contains the specifications of the relevant terms of the
Hamiltonian (i.e., the potential energy and coupling curves), dipole moment
curves as well as %a series of
options determining the method used for the
solution, convergence thresholds etc. Different couplings, corrections or
tasks are switched on by adding the corresponding section %control descriptions
to the input file, i.e. without any alternation of the code.
The input is controlled by keywords and makes use of Stone's input
parser~\cite{input.parser}. All keywords and options are fully documented in
the manual provided along with the source code. The structure of the program
is illustrated in Fig.~\ref{f:program}.

% Along with the program source the library also contains a comprehensive \Duo\ manual,
In addition to the program source code and the manual, we also provide
makefiles for various Fortran~2003 compilers as well as
a set of four examples with sample inputs and corresponding outputs.
These examples comprise (a) a very simple test based on
numerical solution of a single Morse potential, (b) a fit of a single
$^2\Pi$ state to observed energies based on a recent study of the PS molecule \cite{jtPS},
(c) a fit of a single $^3\Sigma^-$ based on a recent study of the PH molecule \cite{jtPS},
and (d) calculation of the spectrum of ScH involving 6 electronic states based
on recent study \cite{jt599}; the output files in this case are given in ExoMol format
\cite{jt548}. These examples only consider low levels of rotational excitation, $J$,
to make them fast to run.

\section{Conclusion}\label{s:conclusion}
\Duo\ is a highly flexible code for solving the nuclear motion problem for
diatomic molecules with non-adiabatically coupled electronic states. It can
simulate pure rotational, ro-vibrational and rovibronic spectra using an
entirely \ai\ input from electronic structure calculations or semi-empirical
data. The latter can also be obtained within \Duo\ by fitting to experimental
data. \Duo\ is currently being further developed and
extensively used to study a number of diatomic
species~\cite{jt563,jt589,jt599,jt618,jt625} as part of the ExoMol project
\cite{jt528}. This project is primarily interested in hot molecules, but
\Duo\ should be equally useful for studying
% the nuclear motion states of
ultracold diatomic molecules.

\section*{Acknowledgements}

This work is supported by ERC Advanced Investigator Project 267219 and by
EPSRC Collaborative Computational Project Q (CCPQ) on Quantum Dynamics. \Duo\
uses the Fortran~90 input parsing module input.f90 supplied by Anthony J.
Stone, which is gratefully acknowledged. We also thank members of the ExoMol
project Laura McKemmish, Andrei Patrascu, Marcus Vasilios, Frances Sinden,
Thomas Fraser, Audra Blissett, Usama Asari, Pawel Jagoda, Laxmi Prajapat and
Maire Gorman for for their help with testing \Duo. AVS would like to thank the
University College London for hospitality during his visits in 2012-13. The
work in MSU was partially supported by the RFBR grant No. 15-03-03302-a.

%\section*{References}

\bibliographystyle{elsarticle-num-names}
%\bibliography{journals_phys,AlO,CaO,jtj,diatomic,methods,linelists,exogen,PN,NaH,avs1,lorenzo,C2,PH3,CO,NaCl}

\end{document}